\documentclass[]{aa}

\usepackage{txfonts}
\usepackage{graphicx}
\usepackage{natbib}
\bibpunct{(}{)}{;}{a}{}{,}
\usepackage{lscape}
\usepackage{pdflscape}
\usepackage {longtable}

\def\ms{m\,s$^{-1}$}
\def\kms{km\,s$^{-1}$}
\def\arcsec{$^{\prime\prime}$}
\def\arcmin{$^{\prime}$}

\def\teff{$T_{\rm eff}$}
\def\logg{$\log g$}
\def\loggf{$\log gf$}
\def\vsini{$v \sin i$}

\title{VSOP: The Variable Star One-shot Project
\thanks{Based on data obtained at the La Silla Observatory, 
European Southern Observatory, under program ID 077.D-0085.}
}
\subtitle{I. Project presentation and first data release}

\author{T. H. Dall\inst{1} \and 
C.~Foellmi\inst{2} \and
J.~Pritchard\inst{3} \and
G.~Lo Curto\inst{4} \and
C.~Allende Prieto\inst{5} \and
H.~Bruntt\inst{6} \and
P.~J.~Amado\inst{7} \and
T.~Arentoft\inst{8} \and
M.~Baes\inst{9} \and
E.~Depagne\inst{4,10} \and
M.~Fernandez\inst{11} \and
V.~Ivanov\inst{4} \and
L.~Koesterke\inst{5} \and
L.~Monaco\inst{4} \and
K.~O'Brien\inst{4} \and
L.~M.~Sarro\inst{12} \and
I.~Saviane\inst{4} \and
J.~Scharwaechter\inst{4} \and
L.~Schmidtobreick\inst{4} \and
O.~Sch\"utz\inst{4} \and
A.~Seifahrt\inst{13} \and
F.~Selman\inst{4} \and
M.~Stefanon\inst{4} \and
M.~Sterzik\inst{4}
}
\institute{Gemini Observatory, 670 N. A'ohoku Pl., Hilo, HI 96720, U.S.A.
\and 
Laboratoire d'AstrOphysique de Grenoble, 414 rue de la Piscine, 38400 Saint-Martin d'H\`eres, France 
\and 
European Southern Observatory, Karl Schwarzschild-Str. 2, D-85748 Garching bei M\"unchen, Germany 
\and 
European Southern Observatory, Alonso de Cordova 3107, Casilla 19001, Vitacura, Santiago, Chile  
\and 
McDonald Observatory and Department of Astronomy, The University of Texas, Austin, TX 78712-1083, U.S.A.
\and 
School of Physics A28, University of Sydney, 2006 NSW, Australia
\and 
Universidad de Granada-IAA(CSIC), P.O. Box 3004, 18080 Granada, Spain
\and 
Department of Physics and Astronomy, University of Aarhus, 8000 Aarhus C, Denmark
\and 
Sterrenkundig Observatorium, Universiteit Gent, Krijgslaan 281 S9, 9000 Gent, Belgium 
\and 
Pontificia Universidad Catolica de Chile, Vicuna Mackenna 4860, Santiago de Chile, Chile
\and 
Instituto de Astrof\'{\i}sica de Andaluc\'{\i}a, Camino Bajo de Hu\'etor 50, 18008 Granada, Spain 
\and 
Department of Artificial Intelligence, ETSI Inform\'atica, Juan del Rosal 16, 28040 Madrid, Spain
\and 
Astrophysikalisches Institut und Universit\"ats-Sternwarte Jena, Schillerg\"asschen 2, 07745 Jena, Germany
}

\offprints{T. Dall  \email{tdall@gemini.edu}}

\abstract
{About 500 new variable stars  enter the General Catalogue of Variable
Stars  (GCVS) every  year.  Most  of them  however  lack spectroscopic
observations, which remains critical  for a correct assignement of the
variability type and for the understanding of the object.}
{The Variable Star  One-shot Project (VSOP) is aimed  at (1) providing
  the  variability type and  spectral type  of all  unstudied variable
  stars,
(2) process, publish, and make  the data available as automatically as
possible,  and  (3) generate  serendipitous  discoveries.  This  first
paper describes the project itself,  the acquisition of the data, the
dataflow, the spectroscopic analysis  and the on-line availability of
the fully calibrated  and reduced data.  We also  present the results
on the 221 stars observed during the first semester of the project.}
{We used the high-resolution  echelle spectrographs HARPS and FEROS in
the ESO La  Silla Observatory (Chile) to survey known variable stars.  Once reduced  by the dedicated
pipelines, the radial velocities are determined from cross correlation
with synthetic template spectra, and the spectral types are determined
by an  automatic minimum distance matching to  synthetic spectra, with
traditional manual spectral typing cross-checks. The variability types
are determined  by manually evaluating the available  light curves and
the spectroscopy. In the future, 
 a new automatic classifier, currently being  developed  by  members  of   the  VSOP  team, 
 based  on  these spectroscopic data and on the photometric classifier developed 
for the COROT and Gaia space missions, will be used.}
{We confirm  or revise spectral types  of 221 variable  stars from the
GCVS.  We identify 26  previously unknown multiple systems, among them
several   visual  binaries   with   spectroscopic  binary   individual
components.   We  present  new  individual results  for  the  multiple
systems  \object{V349 Vel}  and  \object{BC Gru},
for the  composite spectrum  star \object{V4385  Sgr},  for the
T-Tauri  star \object{V1045  Sco}, and  for \object{DM  Boo}  which we
re-classify as a BY Draconis  variable.  The complete data release can
be accessed via the VSOP web site.}


\date{Received $<$date$>$, Accepted $<$date$>$}

\titlerunning{VSOP I. First Data Release}

\begin{document}
\maketitle

\section{Introduction}
\label{intro}
There are more than 38,000 known variable stars listed in the latest edition of the General 
Catalog of Variable Stars \citep[GCVS;][]{kholopov+1998}. Almost 4,000 of these have no 
spectral type assigned and nearly 2,000 are listed with an uncertain variability type, often 
because of lack of spectroscopic characterisation.  The rate of inclusion of new variables is 
currently around 500 per year; actually 1,700 between the Namelists 77 \citep{kazarovets+2003} 
and 78 \citep{kazarovets+2006}. The true number of new variables is higher though, and this 
incompleteness of the GCVS will likely increase in the coming decade due to the large surveys that will
be performed with both ground-based and space-based telescopes.
 About half of the   newly identified GCVS variables have 
unknown variability type and most of them have no published spectral type. Moreover, 
many (even ``firm'') variables have disagreeing designations between different authors 
and even between different catalogs, e.g. there are frequent disagreements between the SIMBAD 
database and GCVS. In addition, binarity is rarely detectable unquestionably by photometric data alone.
Finally, many designations are taken at face value without questioning 
the reliability. This unreliability is a major obstruction to many individual 
studies, and would often require only one ``snapshot'' spectrum to achieve a 
major improvement. Of course, single shot spectra would not always be able to reveal binarity or
transient phenomena.  

Recent examples of the misidentification of variables, where the designation was based solely 
on photometric light curve appearance, and subsequently corrected by taking one single snapshot
spectrum, include:

\object{FH Leo}, that was long thought to be the only known cataclysmic variable (CV) to form 
part of a binary system, being designated as a nova-like variable by \citet{kazarovets+2003} 
based on an outburst observed by the \emph{Hipparcos} satellite. High-resolution FEROS 
spectroscopy allowed us to refute the classification and show that the stars are normal 
F8 and G0 dwarfs \citep{dall+2005}, and that the outburst cannot possibly be due to an accretion 
disk, but rather to a superflare or to erroneous \emph{Hipparcos} measurements, or 
due to a CV ``hidden'' in the light of the two normal stars \citep[][]{vogt2006}.

\object{XY Pic} was included in a study of statistical properties of W UMa type variables (active, 
very fast rotating contact binaries) by \citet{selam2004}, who concluded that it was among the 
most active stars of the sample, based on a fit of its \emph{Hipparcos} light curve, using 
synthetic light curves based on physical parameters. FEROS spectra of XY Pic allowed us to 
show that the star is a rather slowly rotating F3 giant, with no measurable 
chromospheric activity \citep{dall2005}, and is likely a $\delta$~Scuti pulsator. 
This example shows, that even a high-quality  
light curve analysis can result in wrong conclusions about the nature of an object without 
spectroscopic confirmation.

\object{TV Ret} was long thought to be a CV due to an outburst observed 
photometrically in 1977. A single low resolution spectrum, revealed the object to be a compact 
emission line galaxy at $z \sim 0.1$, possibly hosting an extremely bright supernova as the cause of the 
outburst \citep{tvret}.

The above examples illustrate the need for snapshot spectra, and shows that the vast collection 
of poorly studied variable stars contains many errors in terms of variability type designation, which may in many cases 
``cover up'' some potentially interesting physical phenomena under a wrong and seemingly 
dull label. 
In this paper we describe a new large project, the \emph{Variable Stars One-shot 
Project (VSOP)}, undertaken to provide the required ``snapshots''.  We present the
motivation and scope of VSOP in Sect.~\ref{project}, the instrumentation and data handling 
in Sect.~\ref{observations}, and present results 
from the first observing semester from the European Southern Observatory's La Silla site in Sect.~\ref{results},
listing the revised spectral and variability types for 221 stars. 
The results and the reduced data are freely accessible from our website.
We conclude the paper with plans for the future of VSOP in Sect.~\ref{future}, where we also
address the problem of automatic variability classification.

\section{The Project}\label{project}
Motivated by the 
situation outlined above, the
goals of VSOP are:
\begin{enumerate}
\item To obtain the first spectroscopy of all unstudied variable stars, revising spectral and variability types.
\item To process, publish, and make the data available as automatically as possible, facilitating additional science.
\item To generate serendipitous discoveries that will fuel future research.
\end{enumerate}
In addition, due to its beginnings as a proposal for an observatory project, VSOP has been designed with
a view to both science and observatory efficiency.

\subsection{Science efficiency}
A stellar spectrum is a rich source of information. Often, however, only certain aspects of the object
under study is of interest to the scientist conducting the study. One main goal of VSOP is to revise spectral and
variability types, but there are certainly many other scientifically interesting studies one could perform using 
our data. We choose, rather than to sit on the data until we may find time for additional studies, to make
the data easily available to the general community, in the hope that somebody else will be able to do
additional science with the data.  This way, the science output is maximised.

Another aspect that contributes to the science efficiency, is serendipity.  The VSOP observations
are targeting poorly studied variable stars, many of which are exhibiting poorly studied phenomena. We
thus expect to obtain by chance\footnote{''Chance favors the prepared mind`` -- L.~Pasteur} 
data that either merit follow-up in-depth studies, or sheds light on some hitherto obscured phenomenon.
Much of this work may naturally be done by groups not affiliated with VSOP.

\subsection{Observatory efficiency}
VSOP was originally conceived as an ESO observatory project, aimed at providing observations with
loose weather and pointing constraints, with the aim of increasing observing efficiency during periods
when other programmes with stronger constraints on airmass, seeing, and/or 
transparency cannot be carried out. 
Given the all-sky coverage, the loose constraints, and the large scope of the project, VSOP is
an excellent example of a perfect filler programme, which will be extended to other observatories in the near future.

\subsection{VSOP vs other surveys}
The most extensive spectroscopic survey to date, is the Sloan Digital Sky survey \citep[SDSS;][]{york+2000}.
As of
the fifth data release, the SDSS provides more than
800,000 spectra of galaxies, quasars and stars over
nearly 7000 square degrees, mostly in the Northern hemisphere.
The spectroscopy is carried out with multi-object fiber
spectrographs, providing spectral resolution of R=2000
and spectral coverage from 3900--9100\,\AA. The fiber
aperture on the sky is 3\arcsec, so some
contamination from fainter nearby objects is possible.

We cross-correlated the GCVS with the SDSS spectroscopic
list to estimate the possible contribution of the SDSS
towards accurate spectroscopic classification of the
variable stars. The overlap consists of 80 objects (less than 0.3\% of the GCVS) and
given the degree of SDSS completeness we expect some
additional objects, on order of 10, leading us to conclude that
less than 0.5\% of the GCVS have been covered by SDSS. The low number of
stars in common is a consequence both of the faint brightness limit of the SDSS ($g>14$), and of the science goals of
the SDSS, which dictated selection for spectroscopic
follow up primarily for the extragalactic targets, obtaining
spectra of stars only if some fibers remained unused.

However, spectra of a significant number of stars will
be obtained under the Sloan Extension for Galactic
Understanding and Exploration \citep[SEGUE;][]{newberg+2003,rockosi2006} 
which plans to obtain spectra of 240,000 Milky
Way stars over 3500 square degrees with the same
spectrograph. The goal of this survey is to provide
radial velocities and metallicities with typical
accuracies of 10~km\,s$^{-1}$ and 0.3~dex respectively.
SEGUE would thus complement VSOP. However, VSOP  provides superior
spectral resolution and $S/N$ at any given magnitude, since we 
use 2--4\,meter-class telescopes. Plus, VSOP is already producing
and releasing data. 

Another large survey is the RAdial Velocity Experiment \citep[RAVE;][]{steinmetz+2006}, aimed 
at kinematic studies of the local Milky Way environment. While this survey targets a large 
number of stars (24,748 in the first data release), the spectral coverage is limited to the 
IR Ca triplet region, and only at a moderate resolution of 7,500.  Thus, RAVE is likely not 
very useful as a general classification and discovery study.

Of existing surveys, the GAUDI \citep{solano+2005} is the one most similar to VSOP. GAUDI is a
photometric and spectroscopic database of objects that may be observed by the COROT mission, covering
all targets down to $V=9.5$ inside the COROT accessibility zone -- an area on the sky of $10^\circ$ radius.
While GAUDI covers a small area of the sky, looking for stellar variability in all available objects, VSOP 
is targeting known variables all over the sky. Thus, while complimentary, our scope is different.

Furthermore, a number of spectroscopic surveys have been performed in recent years, targeting the variability of
individual types or classes of stars. Examples include surveys for $\beta$ Cep stars \citep{telting+2006},
Hipparcos-selected O-B supergiants \citep{lefever+2007}, $\gamma$ Dor stars \citep{decat+2006}, and 
studies of Ap star oscillations \citep{kurtz+2006}.  While these are all high-resolution spectroscopic studies,
they target a limited subset of stellar types, while the scope of VSOP is all of stellar variability, spanning
the complete HR diagram, including all phenomena. In this respect, VSOP is a unique project, and it is our hope that VSOP 
will also turn out to be a unique resource for researchers of any stellar variability phenomena.

\section{Observations \& Data Handling}
\label{observations}

We present here the results of the first semester of VSOP observations, collected between April and October 2006
with the two high-resolution Echelle spectrographs FEROS and HARPS, of the ESO La Silla Observatory in Chile. 

\subsection{Target selection}
\label{targets}
Our target list is constructed based on a magnitude limited sample of spectroscopically 
unstudied southern variables drawn from the  GCVS composed of stars with unknown or uncertain 
variability types (designations ending in ``:''), in addition to the irregular or ``unsolved'' variables 
(GCVS types *, I, L, S and subclasses). This was complemented with stars having disagreeing 
designations according to SIMBAD, or according to recent literature. Stars were chosen to have a 
magnitude generally brighter than $m_v = 10$ to be easily observable with high-resolution spectrographs
on middle-sized telescopes.

The 221 stars reported here, belong to the bright end of the unstudied variables of the 
GCVS, which is now $\sim$40\% complete (i.e., having reliable, wide wavelength coverage spectroscopy) 
down to $m_v = 10$. The rate of discovery of new variables have been 
relatively constant in recent years, and is very low at the bright end of the distribution. Assuming
similar number of observations for the coming semesters, we can expect to complete the bright end of the 
unstudied variables of the Southern hemisphere in 1--2 additional observing seasons.
Going to fainter magnitudes, the completeness decreases rapidly, reaching a plateau of around 20-25\% fainter than $m_v = 13$,
not including as yet unrecognized variables.

\subsection{FEROS}
\label{feros}
FEROS \citep[Fibre-fed Extended Range Optical
Spectrograph,][]{FEROS1999}, is ESO's high resolution, high efficiency
versatile spectrograph. It is a bench-mounted, thermally controlled,
prism-cross-dispersed \'Echelle spectrograph.  It provides in a single
spectrogram spread over 39 orders almost complete\footnote{The two
spectral ranges $853.4$--$854.1$\,nm and $886.2$--$887.5$\,nm are lost
due to non overlap of the spectral orders.} spectral coverage from
$\sim350$--$920$~nm at a resolving power of 48,000.

The spectrograph is fed by two fibres providing simultaneous spectra
of object plus, in the case of VSOP observations, an empty sky region
for background subtraction. The fibres are illuminated via apertures
of 2.0\arcsec\ on the sky separated by 2.9\arcmin.  A dedicated
pipeline implemented as a MIDAS context provides, in almost real-time,
extracted 1-dimensional, wavelength calibrated spectra.

FEROS Period-77 VSOP observations have been obtained with exposure
times ranging from 180\,sec to 1200\,sec. Given the relaxed observing
constraints, signal-to-noise ($S/N$) ranges from $\sim10$ to $\sim370$ at $V$.
The standard calibration plan, which provides bias, flat-field,
wavelength calibration and spectrophotometric standard star
observations, has been used for this programme.

\subsection{HARPS}
\label{harps}
HARPS \citep[High Accuracy Radial velocity Planet Searcher;][]{HARPS2003} is the ESO  
instrument dedicated to extrasolar planet searches through the radial velocity method. 
Moreover, it has proved to be very efficient as well as a general purpose high resolution 
spectrograph.  HARPS is a fibre fed, E\'chelle cross-dispersed spectrograph, achieving a 
resolution of 110,000, while covering the spectral range from 390~nm to 690~nm in 72 spectroscopic 
orders. The spectrograph is kept under vacuum  and under strict temperature control to increase 
stability. The light is fed to the spectrograph through a fibre with a 1~\arcsec\ aperture on the sky. 
Like FEROS, the fibre link includes two fibres, one for the scientific target and the other for 
sky subtraction. The efficiency of the spectrograph peaks at $\sim 8\%$ (Blaze maximum) at $520$~nm 
and is quite flat between $450$~nm and $690$~nm.

The standard calibration set executed prior to each observing night included bias, flat-fields and 
wavelength calibration. The HARPS data are automatically processed upon acquisition by a dedicated 
pipeline developed by the HARPS consortium and which
provides bias subtraction, order localization, flat fielding, cosmics filtering, order 
extraction \citep[using the Horne technique, ][ assigning lower weights to the pixels away 
from the peak in the spatial profile at any given wavelength]{Horne1986} and radial 
velocity determination through cross correlation of each spectral order with a predefined stellar mask (synthetic spectrum).

HARPS Period-77 VSOP observations have been obtained with exposure times ranging from 90\,sec 
to 1200\,sec, resulting in $S/N$ between 30 and 150, averaging to $\sim 100$ at 550~nm.

\subsection{The VSOP Wiki Database and Data Policy}
\label{wiki}

All the data and basic information about the stars are stored in a wiki-wiki
website located at 
{\tt http://vsop.sc.eso.org}, from where the reduced data of this First Data Release can be freely accessed.
We expect to make incoming 
data freely available through subsequent data releases, with only a few months delay to allow for our inital
data analysis.  Research work benefiting from VSOP data should reference this paper, and include the following acknowledgement:
\begin{quote}
Based on data provided by the VSOP collaboration, through the VSOP wiki database operated at ESO
Chile and ESO Garching.
\end{quote}

For the organization of information, we have chosen the MediaWiki software, developed
for the open and free on-line encyclopedia Wikipedia. This ensure a reliable and extendable website where
all VSOP members can contribute easily from their own daily workplace. This is of growing importance given the distribution of 
VSOP members around the world, as evidenced by the list of affiliations for this paper. 

The MediaWiki software is based on the article/discussion wiki philosophy, which means that to each 
article page there is an associated discussion page. For VSOP we have extended the software to make the 
discussion pages restricted to VSOP members only, while the article pages are reserved for already 
published results, freely accessible to anybody. Thus, each star has a dedicated article page, where 
basic informations (coordinates, magnitude, link to SIMBAD, finding charts, old variability and 
spectral types -- when available) are provided. Also, the  observation details are described as 
well as the analysis, its results, a list of references, catalogues and download links to plots 
of the spectra as well as to all the reduced data products: Cross-Correlation Function (CCF) and 
wavelength calibrated one-dimensional spectrum, all of which are publicly available.

Finally, the MediaWiki software allows the wiki website to be scriptable. We have thus developed 
a VSOP-dedicated software module written in Python which makes the development of scripts dedicated  
to VSOP pages much easier. These robot scripts can then update a large amount of repetitive 
information, or collect the results of given subcategories of stars. Table~\ref{table1} of
this paper, for instance, is automatically produced by one of these scripts. 

\subsection{The VSOP Dataflow}
\label{dataflow}
The VSOP dataflow comprises a collection of different steps that are developed in order to make it as
automatic as possible.  The following is an outline:

\begin{enumerate}
\item A list of targets is build, as described in Sect.~\ref{targets}.
\item When observing time is granted, we automatically generate one wiki-page per target.
\item Observations for each target are defined and submitted to the observatory database.
\item Observations are carried out through the observing semester, and data is
automatically reduced by the instrument pipelines (Sects.~\ref{feros} and~\ref{harps}).
\item Raw and reduced data are automatically transferred to the VSOP machine at ESO Vitacura (Sect.~\ref{wiki}) and the wiki star 
page is updated.
\item Automatically generated plots of spectra and CCF are included in the star 
page (Sect.~\ref{binaries}). 
\item VSOP members receive an email alert that new VSOP observations have been obtained.
\item Analysis is undertaken, and the wiki-pages  are updated if needed.
\end{enumerate}

The dataflow has proved very smooth and efficient throughout the first observing season. For the following
seasons, we have in addition implemented automatic spectral analysis (Sect.~\ref{spectyp}).
However, in 
order to make it completely automatic, one needs to develop an automatic variability classifier, 
incorporating both spectroscopy and the available photometry. As already
mentioned, this is an area where VSOP will play an active future role (Sect.~\ref{future}).

\section{Results}
\label{results}

Table~\ref{table1} lists the 221 stars observed during ESO Period 77: 90 of these were observed with HARPS, 131 with FEROS.

\subsection{Radial velocities and binary status}
\label{binaries}
Radial velocities (RVs) are computed via the Cross Correlation Function (CCF) method, in which a 
template synthetic spectrum with box shaped lines is correlated with the star spectrum to measure the 
Doppler line shift. Details can be found in \citet{baranne96}.
The HARPS online pipeline provides accurate radial velocities ($\approx 1$~\ms) for slowly rotating 
late-F, G, K and early M dwarfs. 
Similarly, FEROS demonstrated a RV accuracy of $\approx 20$~\ms,
the difference with HARPS being mainly due to the mechanical stability of the latter and the different 
choice of light injection in the two instruments.
Correlation with mid to late M type stars is problematic due to the amount of wide molecular features 
in their spectra and to the paucity of narrow metal lines.
Earlier type stars often have higher rotational velocities and weaker metallic lines, 
limiting the accuracy of the radial velocity determination but, at least for stars with metal lines, 
still allowing the computation of a CCF and therefore the estimation of the RV.
In the presence of only few metal lines in the early type stars, the correlation with the G2 template
(the ``earliest'' available for HARPS) will return a CCF with a small contrast (few $\%$). 
Due to the scarcity of narrow metal lines, early-type templates will necessarily 
include strong lines which naturally suffer from asymmetries, which in turn will lower the RV precision.
For the time being, the RV of earlier spectral types is 
estimated by automatic fitting of the core of H$\beta$ with a second-order
polynomial as part of the VASP analysis (cf. Sect.~\ref{spectyp}).
While techniques such as the cross correlation in Fourier space with template spectra obtained from 
observations allows relative accuracies of $\approx 10$~\ms\ in RVs on early type stars \citep{galland05}, 
we use the faster CCF method, as such an accurate RV determination is not our primary concern.

We designate SB2 and SB3 binarity status from the presence of multiple peaks in the CCF, or
via a careful analysis of the spectrum, identifying spectral features belonging to stars of different spectral types. The latter
case is when the stars have widely different spectral types, and the CCF mask only ''sees`` one of the components (see e.g.
V4385 Sgr, Sect.~\ref{v4385sgr}). Since we have only one epoch, the SB1 designation is not used.

Out of the 221 observed targets, we identify 22 new SB2 binaries, several of these as components of wide visual binaries.
In addition we find four new SB3 binaries. However,
the binarity of most of our targets remains undetermined due to our single-epoch approach.
For many of the stars we could not compute a CCF, due to the difficulty to build reliable templates for such peculiar objects.
We postpone accurate determination of the binary status of such stars to a later work.

\begin{table*}[ht]
\label{tab:vasp}
\caption{A sample of results from the automatic analyses. 
We list the names, HD numbers and spectral types from the manual spectral typing.
We give the \teff, \logg\ and [Fe/H] and estimated uncertainties determined by VASP and VWA.
\vsini\ values are calculated with VWA and have uncertainties of 10-20\%.
The first four targets are not VSOP targets, but high S/N HARPS spectra taken from \citet{dall+2006}, which we have used
to calibrate our tools. Full detailed results for all VSOP stars can be found online (see text).}
\centering
\begin{tabular}{lrlr|lcc|lcc}\hline
Star         &         &Spectral&\multicolumn{1}{c|}{\vsini}&\multicolumn{3}{c|}{VASP} & \multicolumn{3}{c}{VWA} \\ 
             &  HD     & type   &[\kms]&\teff\ [K]&\logg &[Fe/H]& \teff\ [K]&\logg    & [Fe/H]     \\ \hline
$\alpha$ Hor &  26967  & K1III  &  2 & 4675(143)  & 2.8(4)   &  $+0.22(13)$  & 4550(80)  & 2.2(2)  & $-0.02(5)$ \\ 
$\tau$ Cet   &  10700  & G8V    &  3 & 5292(157)  & 4.6(4)   &  $-0.45(15)$  & 5320(50)  & 4.5(1)  & $-0.47(5)$ \\
Sun          &    $-$  & G2V    &  2 & 5842(168)  & 4.5(4)   &  $-0.16(16)$  & 5810(40)  & 4.5(1)  & $-0.07(3)$ \\
$\gamma$ Ser & 142860  & F6IV   & 11 & 6322(191)  & 4.1(7)   &  $-0.18(20)$  & 6250(80)  & 4.1(1)  & $-0.24(5)$ \\ \hline\hline
ZZ Pyx       &    $-$  & K3V    &  5 & 4228(223)  & 1.4(1)   &  $-0.31(13)$  & 3900(150) & 1.4(3)  & $-1.3(2)$  \\ 
V349 Vel     &  91021  & F2+?   & 16 & 7314(241)  & 5.2(9)   &  $-0.89(54)$  & 7200(150) & 4.0(2)  & $-0.43(5)$ \\
LP Vir       & 115466  & F1IV   & 40 & 7031(194)  & 3.4(8)   &  $-0.32(29)$  & 7030(80)  & 3.7(1)  & $-0.13(5)$ \\
V976 Cen     & 118551  & F0III  & 40 & 7048(203)  & 2.0(7)   &  $-0.16(23)$  & 8400(100) & 2.4(2)  & $-0.78(5)$ \\
PP Hya       &  87130  & A5V    & 80 & 7924(219)  & 4.0(8)   &  $-1.35(73)$  & $-$       & $-$     & $-$        \\
\hline
\end{tabular}
\end{table*}

\subsection{Spectral classification}
\label{spectyp}
The first set of atmospheric parameters ($T_\mathrm{eff}$, $\log g$, [Fe/H]) listed in Table~\ref{tab:vasp} 
have been obtained automatically by comparison
of the observed spectra with theoretical spectra in a region around H$\beta$ (486~nm). 
This spectral window can be used to uniquely constrain the atmospheric parameters for stars 
with spectral types A to K \citep{allende2003}. This automated software (VASP: VSOP Automatic Stellar Parameters) 
and the synthetic spectra are very similar to those used for the 
STELLA robotic telescopes (Strassmeier et al.~2007; in preparation).

The search for the optimal solution is carried out using the Nelder-Mead algorithm and 
third order interpolation in a grid of synthetic spectra based on \citet{kurucz2006} model 
atmospheres and modern line and  continuous opacities\footnote{Also \citet{kurucz1993} and 
{\tt http://kurucz.harvard.edu/}, the odfnew grid.} \citep{allende+2003a,allende+2003b}. 
The grid currently in use covers $4500 < T_\mathrm{eff} < 7500$~K.
To overcome problems with fast rotation, the grid has been constructed 
with a resolution of 38~\kms, corresponding to a spectral resolving power of 7,700. 
For faster rotational velocities, the accuracy 
of the fits decrease significantly as rotation increases.  

The solar reference abundances are from the photospheric values
compiled by \citet{asplund+2005}. Known spectroscopic binaries 
(cf.~Sect.~\ref{binaries}), as well as stars clearly
outside the grid boundaries, are not run through VASP.

Future upgrades to VASP will include wider temperature range grids, 
as currently only about one third of our targets fall within the limits of the grid. Other upgrades will be
parameter estimation using other 
spectral intervals besides H$\beta$, and the ability to handle 
rotational broadening.

For comparison, we have included results from a classical abundance analysis 
obtained with the VWA\footnote{VWA is available here: {\tt http://www.hans.bruntt.dk/vwa/}.} package \citep[][]{bruntt+2004}.
VWA works with the original (full resolution) spectra and determines the abundance of each individual line.
It relies on the lines of Fe, Cr and Ti to automatically adjust the microturbulence, \teff\ and \logg\ of the
applied atmospheric models \citep[][]{heiter+2004}.
VWA is a semi-automatic procedure and to obtain optimal results the user needs to make 
(1) a careful correction of the continuum and 
(2) inspect the fit of individual lines. 
The abundances found with VWA are based on corrected \loggf\ values, 
which are derived from the HARPS spectrum of the Sun. 
We did not analyse PP~Hya due to its high \vsini, which is known to
cause problems for VWA's automatic procedures.
While VWA seems to produce more robust results, the process involves a lot of manual intervention and is
not at this point suited for an automated analysis.

We have in addition performed manual spectral
classifications, by comparison with standard 
stars of the MK spectral classification as defined by \citet{morgan+1978} and \citet{keenan+mcneil1976}. 
The practical comparison has been done with the help of the Digital Spectral Classification Atlas by 
R.\,O.\,Gray\footnote{ \small {\tt http://nedwww.ipac.caltech.edu/level5/Gray/frames.html}}, 
using high resolution spectra of spectral standards obtained with HARPS, FEROS and UVES. 
Whenever the emission cores of the \ion{Ca}{ii} H\&K lines have been present, we have determined the
luminosity type from the Wilson-Bappu effect \citep{wilson+bappu1957}, using the calibration by \citet{pace+2003}.
Since many older classifications are done in this way, it is instructive to investigate the differences between this traditional 
human skill driven task, and a modern automatic classification. 

We have identified several causes for problems in the spectral classifications, the most common one associated with binarity.
Spectroscopic and very close visual binaries (separations $<1.5$\arcsec) often show multiple peaks in the CCF, and are thus easy
to filter out of the VASP analysis. One such example is \object{V349 Vel} (Sect.~\ref{v349vel}). 
We also found that early-type and very metal-weak stars limits the precision of the VASP fit, and of course influences also the
manual classification. 
In many cases, apparent low metallicity may be due to a binary companion contributing light to the spectrum, causing
the metallic lines to appear weaker. One clear example of this can be seen in V4385~Sgr (Sect.~\ref{v4385sgr}).
In a few cases we have found disagreement between the VASP-computed $\log g$ and the absolute magnitude derived from the 
Wilson-Bappu relation, in most cases caused by being near the lower temperature 
limit of the grid.

\subsection{Selected individual cases}
In this section, we give a few examples of different kinds of objects we have encountered. These examples include both
``typical cases'' and ``special cases'', in order to give an impression of
the wide range of phenomena we have to deal with, and which our automated procedures will have to be able to handle. 
Subsequent publications will deal in depth with interesting special cases, be devoted
to resolving binaries, deal with various classes of stars, and will address the general problem of variability classification.

\subsubsection{V349 Vel -- visual and spectroscopic binary}
\label{v349vel}
\begin{figure}
\includegraphics[width=0.9\linewidth]{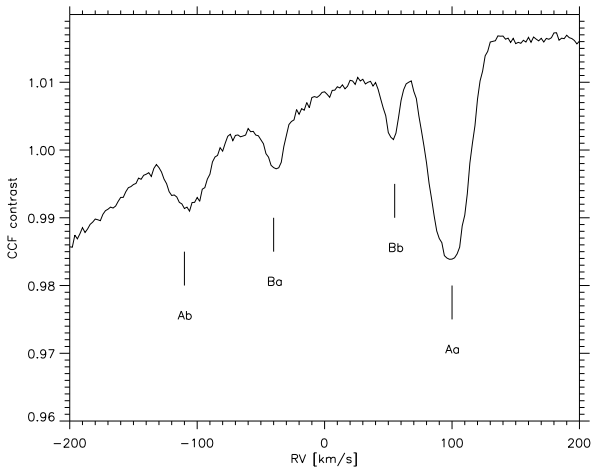}
\caption{\label{fig:v349vel_ccf} The CCF of V349 Vel. Capitals denote the CCDM designation, while lower case letters
label the individual components of each visual component.
}
\end{figure}

\object{V349 Vel} is a close visual binary, separated by $1.1$\arcsec. The CCDM \citep{ccdm1994} lists spectral type F5 for the
A component ($m_v=9.8$), while the B component must be a later type at $m_v=11.1$.  
Our HARPS spectrum, obtained on 2006-04-01, reveals a complex CCF with at least four well defined components, the dominant
one being a F2 type.
Due to their proximity, the fiber entrance (1.0\arcsec) includes light from both visual components. Thus, the A and B
components both spilt up into SB2 spectroscopic binaries. Based on the line strengths and the rotational broadening, which is
expected to be higher for F type stars than for later spectral types, we identify the lines of each of the four components
as shown in Fig.~\ref{fig:v349vel_ccf}.
While new epoch spectroscopy would be needed to determine the orbital parameters, we consider it beyond doubt that V349 Vel
is a four component multiple system composed of two SB2 binaries orbiting each other 

The GCVS variability type is $\alpha^2$~Canum Venaticorum (ACV), 
which  is reserved for magnetic B-A stars with peculiar spectra. 
We suspect that the SB4 appearance of V349 Vel may have been misinterpreted as spectral peculiarities. We do not see
evidence of magnetic activity in the spectrum, hence we also consider BY Dra
type variability unlikely. 
From our spectral analysis (Sect.~\ref{spectyp}) we
find that the parameters of the primary component is 
consistent with an early F~type star near the main sequence.
Thus, the primary component of V349~Vel could 
well belong to the $\delta$~Scuti or $\gamma$~Dor variables. Alternatively,
one or both of the SB2 components could show eclipses.
\citet{otero2003} listed V349 Vel as an EA-type eclipsing binary, with $P = 3.02$~d, and notes that additional
shorter periods may be present as well.
More detailed photometric or spectroscopic observations are needed to 
understand the components of the system better.

\subsubsection{BC Gru -- a triplet eclipsing binary}
\object{BC Gru} is an very short period ($P=0.2662$~d) 
eclipsing contact binary, listed as a late-type contact W system in the catalog
by \citet{malkov+2006}. Apparently, this object has not even been observed with low-resolution or objective prism surveys. 

\begin{figure}
\includegraphics[width=0.9\linewidth]{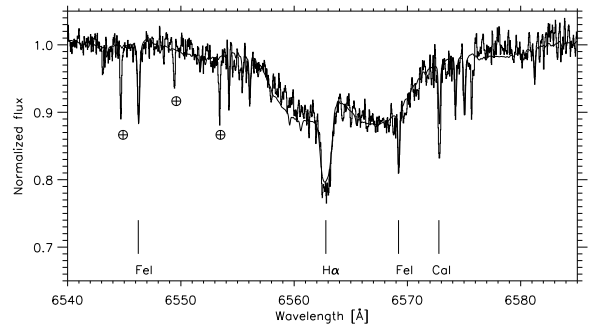}
\caption{\label{fig:bcgru}H$\alpha$ line region of BC Gru. Over-plotted is the best fitting synthetic
spectrum (see text).  The marked lines all belong to the tertiary component, except the telluric lines marked $\oplus$. 
}
\end{figure}

Our FEROS spectrum of BC Gru not only confirms the contact binary nature, but also reveals a third component,
as evident in Fig.~\ref{fig:bcgru}. All three stars have approximately the same spectral type.
We have performed a simple spectral fitting using archive HARPS spectra of the K2V star \object{HD 22049} ($\epsilon$~Eri).
We have artificially broadened and Doppler-shifted two copies of this template, then combined with an unbroadened copy of
the same template to emulate the spectrum of BC~Gru. The fitting has been done using STARMOD \citep{barden1985,montes+1995,montes+2000}. 
A better fit could likely be obtained using
different templates for each component, but in order to do a more accurate analysis, one would need spectra at several different 
orbital phases. Our analysis yields rotational velocities of $(v\sin i)_a = 165\pm50$~\kms\ and $(v\sin i)_b = 142\pm50$~\kms,
while the sharp-lined component is a very slow rotator, with measured $(v\sin i)_c = 6\pm3$~\kms. In Fig.~\ref{fig:bcgru} the $a$
component is the red-most one.

\subsubsection{V4385 Sgr -- composite spectrum}\label{v4385sgr}
\object{V4385 Sgr} is a known composite spectrum star, classified as 
an ellipsoidal variable, and hence a close early-type binary system with 
geometrically distorted components \citep{kazarovets+99}.
Its variability was detected by Hipparcos, which found a period of 2.62~d with peak-to-peak variations
of 0.08~mag. 
It was observed during lunar occultations
\citep[][]{dunham+74,schmidtke+89} and  recorded as double in the
University of Texas  Special Double Star list \citep[][]{schmidtke+79}.
Previous prism observations (typically 39/91~\AA~mm$^{-1}$) 
were obtained by several authors  \citep[see,
e.g.,][]{houk+75,reed+95,garrison+77,abt+79,kennedy+83,kawabata+00}. The
composite spectrum is usually classified combining an
early B to late A star component, plus an early F.
In particular, \citet[][]{garrison+77} classified V~4385~Sgr as a shell star
having Mg, He, and Si like a B5 star but with very broad H and \ion{Ca}{ii} and the rest of
the spectrum like an F2 star. 
SIMBAD however, classifies it as a A2/A3V primary and unknown secondary, which is clearly incorrect.

Our high quality VSOP spectrum of V4385~Sgr
($S/N = 130$ at 550~nm) was obtained on 2006-05-04 using HARPS, and is the 
first full optical range high resolution spectrum of the source.
The HARPS CCF is single-peaked at high contrast (11\%) but asymmetric, as expected due to geometric distortion.
The CCF is constructed using a G2 mask, and hence represents the average line profile of 
the F-component only at a radial velocity of $14.5$~\kms.
The FWHM of the CCF is 13~km\,s$^{-1}$, which is a safe upper limit for $v\sin i$. 
Thus the rotation is significantly slower than
for the average early F-star, which is surprising given the short variability period, and the fact 
that binary evolution tends to increase rotation rates as the orbit
shrinks. 

\begin{figure}
\includegraphics[width=0.9\linewidth]{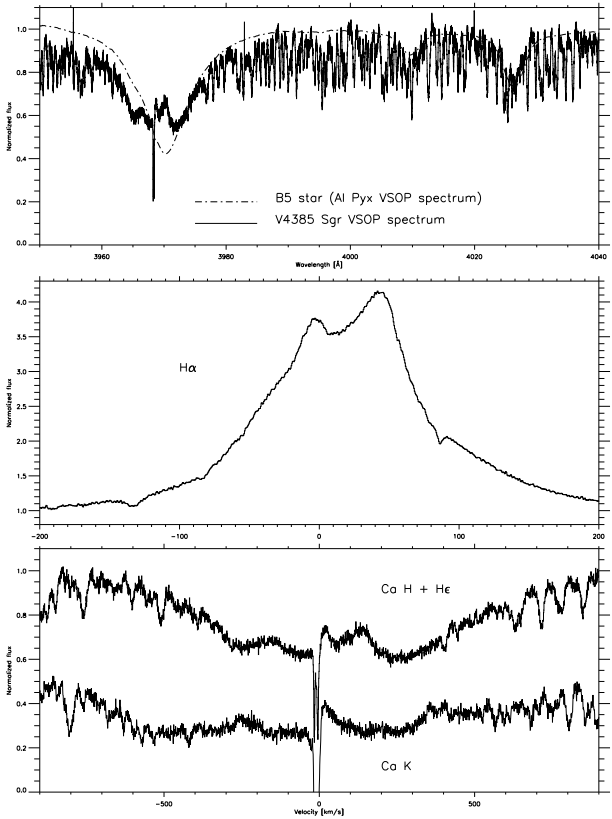}
\caption{\label{fig:v4385sgr}Plot of selected spectral regions of V4385 Sgr.  Fluxes are normalized to the continuum.
\emph{Top:} Over-plotted is the VSOP spectrum of the B5 star \object{AI Pyx}. The composite B+F nature of the V4385 Sgr
spectrum is evident.
\emph{Middle:} H$\alpha$ profile.
\emph{Bottom:} \ion{Ca}{ii} H+K lines. The K line has been shifted down for display
purposes. Note that B stars do not show these lines, so the complex structure (also visible in top panel) comes from
the F-star alone. 
}
\end{figure}

The spectrum (Fig.~\ref{fig:v4385sgr}) 
shows the typical \ion{He}{i} lines of the B component superimposed on the metallic-line spectrum.  Note that
the RV of the two components seem compatible within the uncertainties.

The VSOP spectrum shows Balmer emission lines, strongest at H$\alpha$ (FWHM = 132~km\,s$^{-1}$),
core fill-in at H$\beta$ and H$\gamma$ \citep[see also][]{merrill+50}. 
Weak P~Cyg profiles
can be seen in several lines, however not in the Balmer lines.
Note that the lines of both components, including the strong H$\alpha$ emission, are compatible with
$\mathrm{RV} = 14.5$~\kms. 
The interstellar Na doublet is saturated and presents also a P~Cyg profile from the underlying stellar spectrum. 

Based on the observed emission properties, and on the undisputable two-component nature, we can safely  
rule out that the 2.62~d period can be due to $\gamma$-Dor like pulsations in a single F star. Rather,
due to the slow rotation of the F-star and the apparent same RV of the two components, we propose
that V4385~Sgr could be a near pole-on viewed close binary system, showing only slight eclipses. We propose that
the orbital period is equal to  the photometric period, caused either by a partial eclipse of the smaller component, or by
variations in the wind structure over the orbit. 
Unfortunately, only scattered photometric data exists, as summarized by \citet{reed1998}, making it difficult at
this point to test the hypothesis. Given its brightness,
the star would be an easy target for small telescopes.

\subsubsection{V1045 Sco -- strong lithium absorption} 
\object{V1045 Sco} (HD 144377, V=8.06) is classified as K5III in SIMBAD, with only one 
bibliographic  reference \citep{kazarovets+1999}. It is an IRAS source with a  flux of 2~Jy in 
the 12$\mu$m channel, and decreasing fluxes towards the longer  wavelength bands. The mid-IR 
flux is therefore well above the expected  photospheric level, and indicates circumstellar, dusty, material. 

\begin{figure}
\includegraphics[width=0.9\linewidth]{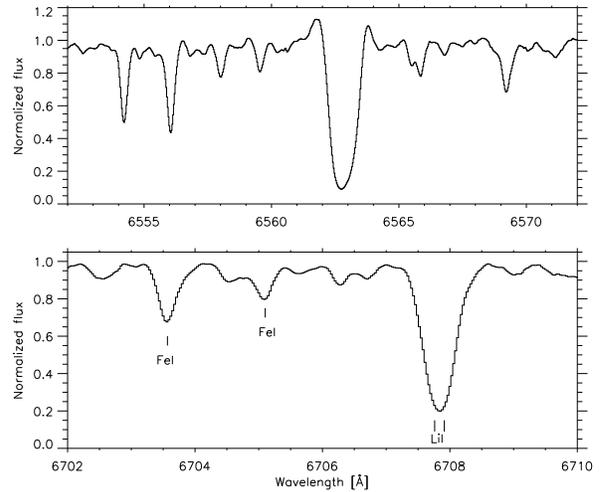}
\caption{\label{fig:v1045sco}Selected spectral regions of V1045 Sco: 
\emph{Top:} H$\alpha$ profile. Note the emission on both the red and blue edges, and the strong reversal. 
\emph{Bottom:} The lithium line region.}
\end{figure}

We have secured two FEROS spectra of this object, on 2006-08-11 and 2006-09-16. Our  spectral 
typing procedure (Sect.~\ref{spectyp}) yields consistently a somewhat later  spectral type, K7, 
compatible with a giant (or sub-giant). A radial velocity of 25.77~\kms\ is determined. 
A striking feature is the H$\alpha$ profile (Fig.~\ref{fig:v1045sco}).
The line is in absorption, but exhibits slightly asymmetric double-peaked emission features in both wings, 
possibly indicative of a rotating disk.

The star exhibits a strong lithium (6708~\AA) absorption, with a measured equivalent width of 0.48~\AA. 
Taken together with its late spectral type, this is a strong youth indicator. Note that the star 
is compatible with being a (weak-lined, as H$\alpha$ emission is inversed) T-Tauri star according 
to the commonly accepted spectroscopic criteria by \citet{martin1997}. Comparing its lithium 
strength to recent measurements in a sample of nearby, young objects \citep{torres+2006}, we conclude that it is 
at least as young as the $\approx$10~Myr old $\beta$~Pictoris moving group members. Its apparent 
youth and its location in the northern outskirts of the Upper Scorpius OB association indicates 
that V1045 Sco might be related to, or in the foreground of, this 5~Myr young association \citep{preibisch+zinnecker1999}. 
This is supported by the apparent presence of remnant circumstellar material, inferred by the 
H$\alpha$ profile and mid-IR excess.
Its measured heliocentric $RV = 25.77$~\kms\ is, however, inconsistent with the bulk motion of 
Upper Scorpius members that peak around $-10$~\kms\ 
\citep[see, e.g.,][]{sartori+2003,jilinski+2006}
The star may therefore well be located in the foreground of the association, possibly related to the Gould Belt 
\citep{guillout+1998}.

\subsubsection{DM Boo -- reclassification}
\object{DM Boo} (HD\,120447,  HIP\,67499, BD+11\,2604) has  been classified as an irregular 
variable star of  spectral type G5 (Simbad) and a Johnson $V$ magnitude  and ($B-V$) color  
of 8.68 and 0.67  respectively. The only photometric monitoring found in the literature is 
the one carried out by  the Northern  Sky Variability Survey \citep[NSVS;][]{wozniak+2004}.   
The  average brightness of the star in their no-filter observations (ROTSE) gives a value 
of 8.59 mag, with a scatter of less than 0.08 mag (and precision of  0.01   mag).  Hipparcos   
gives  a  parallax   for  the   star  of $5.80\pm1.25$~mas, corresponding to a distance of $d=172$~pc.

There are also $uvby$ observations (one  set, from 1993) and the V and B values from  Tycho. 
The Tycho-2 Spectral Type Catalog \citep{wright+2003} converts  the  G5   spectral  class   
into  a   temperature  of $T_{\rm{eff}}=5150$~K.  The GCVS lists the star as a IB type, i.e., a  
poorly studied irregular variable of intermediate to late spectral type.

The VSOP FEROS spectrum was taken on 2006-08-12 with an integration time of 600~s. From the CCF 
we derive a $v\sin{i}$ of $\approx65$~km\,s$^{-1}$ and a radial velocity of $-42$~km\,s$^{-1}$.  
The most prominent features are strong \ion{Ca}{ii} H\&K emission cores, which, together with the 
fast rotation, points to a young active star. Analysis of the lithium line region reveals higher 
than solar photospheric lithium abundance, confirming the notion of a relatively young, active star. 
We thus feel confident classifying DM~Boo as a BY Draconis star, noting the low 
photometric  amplitude typical of such stars.

\subsection{Evaluation in terms of spectral type and binarity}

To compute some statistics of the VSOP observations, we must rely on ''old`` values as they were before VSOP observations.
This concerns mostly the spectral type and the binary status. 
The old spectral type is determined automatically by querying SIMBAD. 
To determine the binary status of a target, we query VizieR catalogues with the Multiple and Double stars, 
Spectroscopic, Cataclysmic and Eclipsing binary keywords to see if the star belongs to one of these classes. 
If the star belongs to a catalogue of spectroscopic binaries, or one providing orbits of stars, the star is said to be a binary. 
If the star belongs to some visual binaries or multiple and double stars catalogs, the star is suspected to be a visual binary. 
Finally, if it only belongs to the Tycho's and Hipparcos catalogues, the suspected 
binary status is inferred from the MultFlag parameter of the catalogue. The procedure 
is quite complex, and is described in the wiki website (Binary Status page). We have 
tested various combination of criteria, which seems to converge to a relatively stable 
results. Our statistics are based on these tests.

The main points of the VSOP observations can be summarized as follows:
\begin{itemize}
\item VSOP observations allow to revise more than two third of the spectral types. 
\item Among these revisions about 4\% had no previous spectral type, about 25\% of the revisions implies 
a complete change of spectral class (say from F to G),
about 30\% are changes by more than one subclass (say from F3 to F5), and about 20\% are changes of one subclass.
The remaining approximately 20\% are assignment or re-assignment of luminosity class.
\item There are no clear trends or systematic biases toward revisions of any given spectral class, spectral subtype, or luminosity class.
\item 40 of our targets (about 20\%) appear to be spectroscopic binaries. Of these, 22 are previously unknown SB2 systems,
while we have identified four new SB3 systems. 
\end{itemize}

\section{The future of VSOP}
\label{future}

This paper has presented the first data release from VSOP, covering ESO Period 77.
Observations are ongoing in Periods 78 and 79, and data from these periods will be released
as soon as the periods end.

\subsection{Next steps}

   Future space missions like COROT  and GAIA will provide a wealth of
   data for  variable stars that  will be observed  with unprecedented
   precision  and  sampling  of  their  light  curves.   Nevertheless,
   situations of ambiguity in the determination of the variability type
   can arise. Also,  as in the case of Hipparcos,  it is expected that
   these data will lead to the discovery of a number of new classes of
   variables.  For all  of them, it will be  necessary to refine their
   variability type and position  in the HR diagram with spectroscopic
   data. This  is what VSOP has  been doing with objects  of the GCVS,
   but the sheer number of stars provided by these missions would deem
   this classification  impossible to be determined  by hand. 

For the next steps,  the main aim of VSOP is to
   acquire  and  develop  the  necessary  tools to  provide  good  and
   reliable   spectral  and   variability   classification  of   stars
   automatically from the available data, either the spectrum, the light
   curve, color information, or combinations of these.

\subsection{Automatic variability classification}

   There have been in the past several attempts to mine large archives
   of variable  objects data, mainly based on  photometric time series
   and, sometimes,  on photometric colours  as well. Good  examples in
   the field  of classification,  are the series  of papers  on MACHO,
   OGLE  and  ASAS  data  (for   RR  Lyrae  stars,  for  example,  see
   \citet{Alco03}, \citet{Sosz03} and  \citet{Wils06}, for MACHO, OGLE
   and   the  NSVS   respectively),  where   catalogues   of  distinct
   variability  classes are  compiled according  to  several selection
   criteria   (rules)    and,   in   some    occasions,   with   human
   intervention. Basically,  the problem of  supervised classification
   of variable  objects can be  described as that of  defining general
   boundaries (hard or  fuzzy) in the hyperspace of  the features that
   describe the classes, based on a set of examples of each class.

   The  field of  Machine Learning  and Pattern  Recognition  offers a
   wealth  of alternatives  for defining  more complex,  flexible, and
   general boundaries  than the hyperboxes used in  the compilation of
   catalogues, minimizing  at the same time the  human intervention in
   the classification process. In  this sense, the VSOP automatic
   variability   classifier  will   build  upon   a   previous  effort
   \citep{sarro06,debosscher+2006,classifier}  carried out  during the
   past few  years to i) create  a well defined training  set of bona
   fide variable objects belonging  to the most important and numerous
   classes,  ii) analyse  the most  relevant and  informative features
   that describe these classes and iii) to study and compare different
   approaches to  the task  of classification, from  Bayesian Networks
   \citep{BN}  to  Bayesian  averages  of artificial  neural  networks
   \citep{Neal}     or      SVMs     (Support     Vector     Machines,
   \citeauthor{Vapnik95}  \citeyear{Vapnik95}).  The development
   of this  classifier was  motivated by the  wealth of  data expected
   from the  COROT space mission \citep{corot} and  was thus designed
   to facilitate  the in-depth  analysis of representative  samples of
   these classes as observed by COROT.  It produces probabilistic
   class  assignements  based on  photometric  time series  parameters
   (harmonic amplitudes  of component frequencies,  phase differences,
   amplitude ratios, etc).  The  objective is twofold: i) to generate
   class  specific  object  lists  for  further  analysis  by  COROT's
   Additional  Program  scientists and  ii)  to  detect objects  lying
   outside the  known density  distribution of objects/classes  in the
   parameter   space;  these  objects   can  possibly   represent  new
   astrophysical  scenarios  for   variability. The  classifier
   presented in  \citet{classifier} has now been extended  by the same
   authors to incorporate the photometric colours $B-V$, $V-I$, $J-H$,
   $H-K$  and,  for a  reduced  number  of  classes, also  Str\"omgren
   indices \citep{classifier2}.

The  effort  described in  the  previous  paragraphs  is now  being
   continued  and adapted  as part  of  the GAIA  Data Processing  and
   Analysis  Consortium \citep{gaia}  to  incorporate attributes  that
   will  be  provided  by  GAIA  instruments  such  as  Blue  and  Red
   spectrophotometry or spectroscopy near the Calcium infrared triplet
   \citep{eyer-class}.

The VSOP automatic classifier  is designed as a Virtual Observatory
   compliant   service  capable   of  producing   probabilistic  class
   assignements  for  objects  with   a  wide  variety  of  attributes
   available (from time  series photometry to multi-wavelength spectra
   or photometric colours) and  will thus represent the culmination of
   all the efforts on which it is based.

   The development plan necessarily includes a first stage where
   spectra of at least a representative sample of the COROT training
   set objects have to be obtained in order to allow for the
   incorporation of spectral information to the classifier.  Here is
   where VSOP initially will play a major part by collecting this
   dataset.  Subsequent to this, a study will be conducted in order to
   determine an optimal subset of features providing the best
   classification performance.  Obvious candidate features will be
   line or band fluxes, equivalent widths, ratios, and/or combinations thereof, line
   asymmetry measures, and derived physical parameters (\teff, \logg,
   [Fe/H]). All these will be subject to the statistical feature
   analysis classical in Machine Learning applications.

   One obvious requisite of the classifier will be its capability for
   prediction based on incomplete data.  This can happen if the
   spectral information only covers a fraction of the wavelength range
   used for training or if the resolution is too poor to separate
   several lines.  Based on these specifications, several
   state-of-the-art machine learning algorithms will be applied to the
   training set and the performance assessed according to standard
   figures of merit like overall misclassification rates or the area
   under the receiver operating curve \citep{roc}.

While these efforts are already undergoing in the context of the space missions, the
VSOP team is actively taking part in these new developments towards
complete automatic classification.

\subsection{Long-term vision}

While it is growing rapidly, VSOP is flexible enough to allow us to imagine a long-term future. 
VSOP is a project implemented only by astronomers, using the MediaWiki software. With a simple, 
though large, collection of scripts developed over the year of operation, in addition to the 
already implemented spectrograph's pipelines, VSOP is a quasi-automatic spectrum production 
machine whose results are automatically available through a wiki website. As emphasized before, 
the spectral and variability analysis will also become automatic in the future. The wiki 
provides a scriptable and clean interface which requires minimum human intervention, while 
centralizing all the work done by the team. In this sense, it combines both the advantages 
of an automatically generated content website (homogeneity, reliability, cleanness), with the 
total flexibility for the contributors to customize a specific point, and for the public to 
have access to the data, the information and the history. In that perspective, we could 
imagine an all-automatic wiki-database accepting {\it pipeline-reduced} spectra from any 
observatories in the world, not necessarily dedicated to variable stars. 

This larger vision of complete automation while retaining absolute flexibility, is at the core of the VSOP future.
As a first step in this direction, we have included the released data in Wikimbad ({\tt http://wikimbad.org}) as
well as directly on our own VSOP server.

\subsection{An Open Community}
We would like to stress that VSOP is an open community of scientists. We obtain, process, publish, and provide data at a
growing rate, and to be able to keep up with that, we welcome any scientist from any background, who wants to contribute
to the growth of the project.  By joining the VSOP Team, first access rights to the data is granted, which has already
spurred several follow-up projects, and will continue to do so.  
A major commitment of VSOP is the fast release of data, including fully calibrated reduced spectra. 
Once released, the data will of course be available to the entire community.

We believe in open sharing of information on all levels, and we believe this can be accomplished without 
sacrificing individual scientific ambitions by basing our collaboration on team-work and the drive for fast
scientific turnover.  We would like to conclude with an open invitation to participate in VSOP and its
mission,
either as part of the VSOP team, or independently through the freely accessible data releases.

\begin{acknowledgements}
This research has made extensive use of the Simbad and VizieR databases, and their XML interface, operated at CDS, Strasbourg, Fance. 
We are very grateful to O.~Hainaut for suggesting the project name, and to the referee for constructive comments that improved the paper.
M.F. was supported by the Spanish grants AYA2004-05395 and AYA2004-21521-E. Supported by the Gemini Observatory, which is operated by 
the Association of Universities for Research in Astronomy, Inc., on behalf of the international Gemini partnership of Argentina, 
Australia, Brazil, Canada, Chile, the United Kingdom, and the United States of America.
\end{acknowledgements}

\bibliography{paper_one}

\begin{thebibliography}{76}
\expandafter\ifx\csname natexlab\endcsname\relax\def\natexlab#1{#1}\fi

\bibitem[{{Abt} {et~al.}(1979){Abt}, {Brodzik}, \& {Schaefer}}]{abt+79}
{Abt}, H.~A., {Brodzik}, D., \& {Schaefer}, B. 1979, \pasp, 91, 176

\bibitem[{{Alcock} {et~al.}(2003){Alcock}, {Alves}, {Becker}, {Bennett},
  {Cook}, {Drake}, {Freeman}, {Geha}, {Griest}, {Kov{\'a}cs}, {Lehner},
  {Marshall}, {Minniti}, {Nelson}, {Peterson}, {Popowski}, {Pratt}, {Quinn},
  {Rodgers}, {Stubbs}, {Sutherland}, {Vandehei}, \& {Welch}}]{Alco03}
{Alcock}, C., {Alves}, D.~R., {Becker}, A., {et~al.} 2003, \apj, 598, 597

\bibitem[{{Allende Prieto}(2003)}]{allende2003}
{Allende Prieto}, C. 2003, \mnras, 339, 1111

\bibitem[{{Allende Prieto} {et~al.}(2003{\natexlab{a}}){Allende Prieto},
  {Hubeny}, \& {Lambert}}]{allende+2003b}
{Allende Prieto}, C., {Hubeny}, I., \& {Lambert}, D.~L. 2003{\natexlab{a}},
  \apj, 591, 1192

\bibitem[{{Allende Prieto} {et~al.}(2003{\natexlab{b}}){Allende Prieto},
  {Lambert}, {Hubeny}, \& {Lanz}}]{allende+2003a}
{Allende Prieto}, C., {Lambert}, D.~L., {Hubeny}, I., \& {Lanz}, T.
  2003{\natexlab{b}}, \apjs, 147, 363

\bibitem[{{Asplund} {et~al.}(2005){Asplund}, {Grevesse}, \&
  {Sauval}}]{asplund+2005}
{Asplund}, M., {Grevesse}, N., \& {Sauval}, A.~J. 2005, in ASP Conf. Ser. 336:
  Cosmic Abundances as Records of Stellar Evolution and Nucleosynthesis, ed.
  T.~G. {Barnes}, III \& F.~N. {Bash}, 25--+

\bibitem[{{Baglin} {et~al.}(2000){Baglin}, {Vauclair}, \& {The COROT
  Team}}]{corot}
{Baglin}, A., {Vauclair}, G., \& {The COROT Team}. 2000, Journal of
  Astrophysics and Astronomy, 21, 319

\bibitem[{{Baranne} {et~al.}(1996){Baranne}, {Queloz}, {Mayor}, {Adrianzyk},
  {Knispel}, {Kohler}, {Lacroix}, {Meunier}, {Rimbaud}, \& {Vin}}]{baranne96}
{Baranne}, A., {Queloz}, D., {Mayor}, M., {et~al.} 1996, \aaps, 119, 373

\bibitem[{{Barden}(1985)}]{barden1985}
{Barden}, S.~C. 1985, \apj, 295, 162

\bibitem[{{Bruntt} {et~al.}(2004){Bruntt}, {Bikmaev}, {Catala}, {Solano},
  {Gillon}, {Magain}, {Van't Veer-Menneret}, {St{\"u}tz}, {Weiss}, {Ballereau},
  {Bouret}, {Charpinet}, {Hua}, {Katz}, {Ligni{\`e}res}, \&
  {Lueftinger}}]{bruntt+2004}
{Bruntt}, H., {Bikmaev}, I.~F., {Catala}, C., {et~al.} 2004, \aap, 425, 683

\bibitem[{{Dall}(2005)}]{dall2005}
{Dall}, T.~H. 2005, Informational Bulletin on Variable Stars, 5617, 1

\bibitem[{{Dall} {et~al.}(2006){Dall}, {Santos}, {Arentoft}, {Bedding}, \&
  {Kjeldsen}}]{dall+2006}
{Dall}, T.~H., {Santos}, N.~C., {Arentoft}, T., {Bedding}, T.~R., \&
  {Kjeldsen}, H. 2006, \aap, 454, 341

\bibitem[{{Dall} {et~al.}(2005){Dall}, {Schmidtobreick}, {Santos}, \&
  {Israelian}}]{dall+2005}
{Dall}, T.~H., {Schmidtobreick}, L., {Santos}, N.~C., \& {Israelian}, G. 2005,
  \aap, 438, 317

\bibitem[{{de Cat} {et~al.}(2006){de Cat}, {Eyer}, {Cuypers}, {Aerts},
  {Vandenbussche}, {Uytterhoeven}, {Reyniers}, {Kolenberg}, {Groenewegen},
  {Raskin}, {Maas}, \& {Jankov}}]{decat+2006}
{de Cat}, P., {Eyer}, L., {Cuypers}, J., {et~al.} 2006, \aap, 449, 281

\bibitem[{{Debosscher} {et~al.}(2006){Debosscher}, {Aerts}, \&
  {Vandenbussche}}]{debosscher+2006}
{Debosscher}, J., {Aerts}, C., \& {Vandenbussche}, B. 2006, in Astronomical
  Society of the Pacific Conference Series, Vol. 349, Astrophysics of Variable
  Stars, ed. C.~{Sterken} \& C.~{Aerts}, 219--+

\bibitem[{{Debosscher} {et~al.}(2007){Debosscher}, {Sarro}, {Aerts}, {Cuypers},
  {Vandenbussche}, {Garrido}, \& {Solano}}]{classifier}
{Debosscher}, J., {Sarro}, L., {Aerts}, C., {et~al.} 2007, {\aap} (submitted)

\bibitem[{{Dommanget} \& {Nys}(1994)}]{ccdm1994}
{Dommanget}, J. \& {Nys}, O. 1994, Communications de l'Observatoire Royal de
  Belgique, 115, 1

\bibitem[{{Dunham}(1974)}]{dunham+74}
{Dunham}, D.~W. 1974, Occultation Newsletter, 1, 5

\bibitem[{{Eyer}(2006)}]{eyer-class}
{Eyer}, L. 2006, in Astronomical Society of the Pacific Conference Series, Vol.
  349, Astrophysics of Variable Stars, ed. C.~{Sterken} \& C.~{Aerts}, 15--+

\bibitem[{Fawcett(2003)}]{roc}
Fawcett, T. 2003, ROC Graphs: Notes and Practical Considerations for Data
  Mining Researchers

\bibitem[{{Galland} {et~al.}(2005){Galland}, {Lagrange}, {Udry}, {Chelli},
  {Pepe}, {Queloz}, {Beuzit}, \& {Mayor}}]{galland05}
{Galland}, F., {Lagrange}, A.-M., {Udry}, S., {et~al.} 2005, \aap, 443, 337

\bibitem[{{Garrison} {et~al.}(1977){Garrison}, {Hiltner}, \&
  {Schild}}]{garrison+77}
{Garrison}, R.~F., {Hiltner}, W.~A., \& {Schild}, R.~E. 1977, \apjs, 35, 111

\bibitem[{{Gilmore} {et~al.}(2000){Gilmore}, {de Boer}, {Favata}, {Hoeg},
  {Lattanzi}, {Lindegren}, {Luri}, {Mignard}, {Perryman}, \& {de Zeeuw}}]{gaia}
{Gilmore}, G.~F., {de Boer}, K.~S., {Favata}, F., {et~al.} 2000, in Proc. SPIE
  Vol. 4013, p. 453-472, UV, Optical, and IR Space Telescopes and Instruments,
  James B. Breckinridge; Peter Jakobsen; Eds., ed. J.~B. {Breckinridge} \&
  P.~{Jakobsen}, 453--472

\bibitem[{{Guillout} {et~al.}(1998){Guillout}, {Sterzik}, {Schmitt}, {Motch},
  \& {Neuhaeuser}}]{guillout+1998}
{Guillout}, P., {Sterzik}, M.~F., {Schmitt}, J.~H.~M.~M., {Motch}, C., \&
  {Neuhaeuser}, R. 1998, \aap, 337, 113

\bibitem[{{Heiter} {et~al.}(2002){Heiter}, {Kupka}, {van't Veer-Menneret},
  {Barban}, {Weiss}, {Goupil}, {Schmidt}, {Katz}, \& {Garrido}}]{heiter+2004}
{Heiter}, U., {Kupka}, F., {van't Veer-Menneret}, C., {et~al.} 2002, \aap, 392,
  619

\bibitem[{{Horne}(1986)}]{Horne1986}
{Horne}, K. 1986, \pasp, 98, 609

\bibitem[{{Houk} \& {Cowley}(1975)}]{houk+75}
{Houk}, N. \& {Cowley}, A.~P. 1975, University of Michigan Catalogue of
  Two-Dimensional Spectral Types for the HD Stars (Ann Arbor: Univ. Michigan)

\bibitem[{{Jilinski} {et~al.}(2006){Jilinski}, {Daflon}, {Cunha}, \& {de La
  Reza}}]{jilinski+2006}
{Jilinski}, E., {Daflon}, S., {Cunha}, K., \& {de La Reza}, R. 2006, \aap, 448,
  1001

\bibitem[{{Kaufer} {et~al.}(1999){Kaufer}, {Stahl}, {Tubbesing}, {Norregaard},
  {Avila}, {Francois}, {Pasquini}, \& {Pizzella}}]{FEROS1999}
{Kaufer}, A., {Stahl}, O., {Tubbesing}, S., {et~al.} 1999, The Messenger, 95, 8

\bibitem[{{Kawabata} {et~al.}(2000){Kawabata}, {Hirata}, {Ikeda}, {Akitaya},
  {Seki}, {Matsumura}, \& {Okazaki}}]{kawabata+00}
{Kawabata}, K.~S., {Hirata}, R., {Ikeda}, Y., {et~al.} 2000, \apj, 540, 429

\bibitem[{{Kazarovets} {et~al.}(1999{\natexlab{a}}){Kazarovets}, {Samus},
  {Durlevich}, {Frolov}, {Antipin}, {Kireeva}, \& {Pastukhova}}]{kazarovets+99}
{Kazarovets}, A.~V., {Samus}, N.~N., {Durlevich}, O.~V., {et~al.}
  1999{\natexlab{a}}, Informational Bulletin on Variable Stars, 4659, 1

\bibitem[{{Kazarovets} {et~al.}(1999{\natexlab{b}}){Kazarovets}, {Samus},
  {Durlevich}, {Frolov}, {Antipin}, {Kireeva}, \&
  {Pastukhova}}]{kazarovets+1999}
{Kazarovets}, A.~V., {Samus}, N.~N., {Durlevich}, O.~V., {et~al.}
  1999{\natexlab{b}}, Informational Bulletin on Variable Stars, 4659, 1

\bibitem[{{Kazarovets} {et~al.}(2003){Kazarovets}, {Kireeva}, {Samus}, \&
  {Durlevich}}]{kazarovets+2003}
{Kazarovets}, E.~V., {Kireeva}, N.~N., {Samus}, N.~N., \& {Durlevich}, O.~V.
  2003, Informational Bulletin on Variable Stars, 5422, 1

\bibitem[{{Kazarovets} {et~al.}(2006){Kazarovets}, {Samus}, {Durlevich},
  {Kireeva}, \& {Pastukhova}}]{kazarovets+2006}
{Kazarovets}, E.~V., {Samus}, N.~N., {Durlevich}, O.~V., {Kireeva}, N.~N., \&
  {Pastukhova}, E.~N. 2006, Informational Bulletin on Variable Stars, 5721, 1

\bibitem[{{Keenan} \& {McNeil}(1976)}]{keenan+mcneil1976}
{Keenan}, P.~C. \& {McNeil}, R.~C. 1976, {An atlas of spectra of the cooler
  stars: Types G,K,M,S, and C. Part 1: Introduction and tables} (Columbus: Ohio
  State University Press, 1976)

\bibitem[{{Kennedy}(1983)}]{kennedy+83}
{Kennedy}, P.~M. 1983, MK Classification Catalogue Extension (Weston Creek: Mt.
  Stromlo Obs.)

\bibitem[{{Kholopov} {et~al.}(1998){Kholopov}, {Samus}, {Frolov}, {Goranskij},
  {Gorynya}, {Karitskaya}, {Kazarovets}, {Kireeva}, {Kukarkina}, {Kurochkin},
  {Medvedeva}, {Pastukhova}, {Perova}, {Rastorguev}, \&
  {Shugarov}}]{kholopov+1998}
{Kholopov}, P.~N., {Samus}, N.~N., {Frolov}, M.~S., {et~al.} 1998, in Combined
  General Catalogue of Variable Stars, 4.1 Ed (II/214A). (1998)

\bibitem[{{Kurtz} {et~al.}(2006){Kurtz}, {Elkin}, \& {Mathys}}]{kurtz+2006}
{Kurtz}, D.~W., {Elkin}, V.~G., \& {Mathys}, G. 2006, \mnras, 370, 1274

\bibitem[{{Kurucz}(1993)}]{kurucz1993}
{Kurucz}, R. 1993, ATLAS9 Stellar Atmosphere Programs and 2 km/s grid.~Kurucz
  CD-ROM No.~13.~ Cambridge, Mass.: Smithsonian Astrophysical Observatory

\bibitem[{{Kurucz}(2006)}]{kurucz2006}
{Kurucz}, R.~L. 2006, in EAS Publications Series, ed. P.~{Stee}, 129--155

\bibitem[{{Lefever} {et~al.}(2007){Lefever}, {Puls}, \& {Aerts}}]{lefever+2007}
{Lefever}, K., {Puls}, J., \& {Aerts}, C. 2007, \aap, 463, 1093

\bibitem[{{Malkov} {et~al.}(2006){Malkov}, {Oblak}, {Snegireva}, \&
  {Torra}}]{malkov+2006}
{Malkov}, O.~Y., {Oblak}, E., {Snegireva}, E.~A., \& {Torra}, J. 2006, \aap,
  446, 785

\bibitem[{{Martin}(1997)}]{martin1997}
{Martin}, E.~L. 1997, \aap, 321, 492

\bibitem[{{Mayor} {et~al.}(2003){Mayor}, {Pepe}, {Queloz}, {Bouchy},
  {Rupprecht}, {Lo Curto}, {Avila}, {Benz}, {Bertaux}, {Bonfils}, {dall},
  {Dekker}, {Delabre}, {Eckert}, {Fleury}, {Gilliotte}, {Gojak}, {Guzman},
  {Kohler}, {Lizon}, {Longinotti}, {Lovis}, {Megevand}, {Pasquini}, {Reyes},
  {Sivan}, {Sosnowska}, {Soto}, {Udry}, {van Kesteren}, {Weber}, \&
  {Weilenmann}}]{HARPS2003}
{Mayor}, M., {Pepe}, F., {Queloz}, D., {et~al.} 2003, The Messenger, 114, 20

\bibitem[{{Merrill} \& {Burwell}(1950)}]{merrill+50}
{Merrill}, P.~W. \& {Burwell}, C.~G. 1950, \apj, 112, 72

\bibitem[{{Montes} {et~al.}(1995){Montes}, {de Castro}, {Fernandez-Figueroa},
  \& {Cornide}}]{montes+1995}
{Montes}, D., {de Castro}, E., {Fernandez-Figueroa}, M.~J., \& {Cornide}, M.
  1995, \aaps, 114, 287

\bibitem[{{Montes} {et~al.}(2000){Montes}, {Fern{\'a}ndez-Figueroa}, {De
  Castro}, {Cornide}, {Latorre}, \& {Sanz-Forcada}}]{montes+2000}
{Montes}, D., {Fern{\'a}ndez-Figueroa}, M.~J., {De Castro}, E., {et~al.} 2000,
  \aaps, 146, 103

\bibitem[{{Morgan} {et~al.}(1978){Morgan}, {Abt}, \& {Tapscott}}]{morgan+1978}
{Morgan}, W.~W., {Abt}, H.~A., \& {Tapscott}, J.~W. 1978, {Revised MK Spectral
  Atlas for stars earlier than the sun} (Williams Bay: Yerkes Observatory, and
  Tucson: Kitt Peak National Observatory, 1978)

\bibitem[{Neal(1996)}]{Neal}
Neal, R.~M. 1996, Bayesian Learning for Neural Networks (New York: Lecture
  Notes in StatisticsSpringer Verlag)

\bibitem[{{Newberg} \& {Sloan Digital Sky Survey
  Collaboration}(2003)}]{newberg+2003}
{Newberg}, H.~J. \& {Sloan Digital Sky Survey Collaboration}. 2003, in Bulletin
  of the American Astronomical Society, 1385--+

\bibitem[{{Otero}(2003)}]{otero2003}
{Otero}, S.~A. 2003, Informational Bulletin on Variable Stars, 5480, 1

\bibitem[{{Pace} {et~al.}(2003){Pace}, {Pasquini}, \& {Ortolani}}]{pace+2003}
{Pace}, G., {Pasquini}, L., \& {Ortolani}, S. 2003, \aap, 401, 997

\bibitem[{Pearl(1988)}]{BN}
Pearl, J. 1988, Probabilistic reasoning in intelligent systems: networks of
  plausible inference (San Francisco, CA, USA: Morgan Kaufmann Publishers Inc.)

\bibitem[{{Preibisch} \& {Zinnecker}(1999)}]{preibisch+zinnecker1999}
{Preibisch}, T. \& {Zinnecker}, H. 1999, \aj, 117, 2381

\bibitem[{{Reed}(1998)}]{reed1998}
{Reed}, B.~C. 1998, \apjs, 115, 271

\bibitem[{{Reed} \& {Beatty}(1995)}]{reed+95}
{Reed}, B.~C. \& {Beatty}, A.~E. 1995, \apjs, 97, 189

\bibitem[{{Rockosi}(2006)}]{rockosi2006}
{Rockosi}, C.~M. 2006, in American Astronomical Society Meeting Abstracts,
  172.04--+

\bibitem[{{Sarro} {et~al.}(2007){Sarro}, {Debosscher}, {L\'opez}, \&
  {Aerts}}]{classifier2}
{Sarro}, L., {Debosscher}, J., {L\'opez}, M., \& {Aerts}, C. 2007, {\aap} (in
  preparation)

\bibitem[{{Sarro} {et~al.}(2006){Sarro}, {Debosscher}, {Aerts}, {Garrido}, \&
  {Vandenbussche}}]{sarro06}
{Sarro}, L.~M., {Debosscher}, J., {Aerts}, C., {Garrido}, R.~{Solano}, E., \&
  {Vandenbussche}, B. 2006, in ''The CoRoT Mission'', (Eds) M. Fridlund, A.
  Baglin, J. Lochard \& L. Conroy, ESA Publications Division, ESA Spec.Publ.
  1306, 385--391

\bibitem[{{Sartori} {et~al.}(2003){Sartori}, {L{\'e}pine}, \&
  {Dias}}]{sartori+2003}
{Sartori}, M.~J., {L{\'e}pine}, J.~R.~D., \& {Dias}, W.~S. 2003, \aap, 404, 913

\bibitem[{{Schmidtke}(1979)}]{schmidtke+79}
{Schmidtke}, P.~C. 1979, \pasp, 91, 674

\bibitem[{{Schmidtke} {et~al.}(1989){Schmidtke}, {Africano}, \&
  {Quigley}}]{schmidtke+89}
{Schmidtke}, P.~C., {Africano}, J.~L., \& {Quigley}, R. 1989, \aj, 97, 909

\bibitem[{{Schmidtobreick} {et~al.}(2007){Schmidtobreick}, {Tappert}, {Horst},
  {Saviane}, \& {Lidman}}]{tvret}
{Schmidtobreick}, L., {Tappert}, C., {Horst}, H., {Saviane}, I., \& {Lidman},
  C. 2007, \aap, 461, 943

\bibitem[{{Selam}(2004)}]{selam2004}
{Selam}, S.~O. 2004, \aap, 416, 1097

\bibitem[{{Solano} {et~al.}(2005){Solano}, {Catala}, {Garrido}, {Poretti},
  {Janot-Pacheco}, {Guti{\'e}rrez}, {Gonz{\'a}lez}, {Mantegazza}, {Neiner},
  {Fremat}, {Charpinet}, {Weiss}, {Amado}, {Rainer}, {Tsymbal}, {Lyashko},
  {Ballereau}, {Bouret}, {Hua}, {Katz}, {Ligni{\`e}res}, {L{\"u}ftinger},
  {Mittermayer}, {Nesvacil}, {Soubiran}, {van't Veer-Menneret}, {Goupil},
  {Costa}, {Rolland}, {Antonello}, {Bossi}, {Buzzoni}, {Rodrigo}, {Aerts},
  {Butler}, {Guenther}, \& {Hatzes}}]{solano+2005}
{Solano}, E., {Catala}, C., {Garrido}, R., {et~al.} 2005, \aj, 129, 547

\bibitem[{{Soszynski} {et~al.}(2003){Soszynski}, {Udalski}, {Szymanski},
  {Kubiak}, {Pietrzynski}, {Wozniak}, {Zebrun}, {Szewczyk}, \&
  {Wyrzykowski}}]{Sosz03}
{Soszynski}, I., {Udalski}, A., {Szymanski}, M., {et~al.} 2003, Acta
  Astronomica, 53, 93

\bibitem[{{Steinmetz} {et~al.}(2006){Steinmetz}, {Zwitter}, {Siebert},
  {Watson}, {Freeman}, {Munari}, {Campbell}, {Williams}, {Seabroke}, {Wyse},
  {Parker}, {Bienaym{\'e}}, {Roeser}, {Gibson}, {Gilmore}, {Grebel}, {Helmi},
  {Navarro}, {Burton}, {Cass}, {Dawe}, {Fiegert}, {Hartley}, {Russell},
  {Saunders}, {Enke}, {Bailin}, {Binney}, {Bland-Hawthorn}, {Boeche}, {Dehnen},
  {Eisenstein}, {Evans}, {Fiorucci}, {Fulbright}, {Gerhard}, {Jauregi}, {Kelz},
  {Mijovi{\'c}}, {Minchev}, {Parmentier}, {Pe{\~n}arrubia}, {Quillen}, {Read},
  {Ruchti}, {Scholz}, {Siviero}, {Smith}, {Sordo}, {Veltz}, {Vidrih}, {von
  Berlepsch}, {Boyle}, \& {Schilbach}}]{steinmetz+2006}
{Steinmetz}, M., {Zwitter}, T., {Siebert}, A., {et~al.} 2006, \aj, 132, 1645

\bibitem[{{Telting} {et~al.}(2006){Telting}, {Schrijvers}, {Ilyin},
  {Uytterhoeven}, {de Ridder}, {Aerts}, \& {Henrichs}}]{telting+2006}
{Telting}, J.~H., {Schrijvers}, C., {Ilyin}, I.~V., {et~al.} 2006, \aap, 452,
  945

\bibitem[{{Torres} {et~al.}(2006){Torres}, {Quast}, {da Silva}, {de la Reza},
  {Melo}, \& {Sterzik}}]{torres+2006}
{Torres}, C.~A.~O., {Quast}, G.~R., {da Silva}, L., {et~al.} 2006, ArXiv
  Astrophysics e-prints

\bibitem[{Vapnik(1995)}]{Vapnik95}
Vapnik, V.~N. 1995, The nature of statistical learning theory (New York, NY,
  USA: Springer-Verlag New York, Inc.)

\bibitem[{{Vogt}(2006)}]{vogt2006}
{Vogt}, N. 2006, \aap, 452, 985

\bibitem[{{Wils} {et~al.}(2006){Wils}, {Lloyd}, \& {Bernhard}}]{Wils06}
{Wils}, P., {Lloyd}, C., \& {Bernhard}, K. 2006, \mnras, 368, 1757

\bibitem[{{Wilson} \& {Vainu Bappu}(1957)}]{wilson+bappu1957}
{Wilson}, O.~C. \& {Vainu Bappu}, M.~K. 1957, \apj, 125, 661

\bibitem[{{Wo{\'z}niak} {et~al.}(2004){Wo{\'z}niak}, {Vestrand}, {Akerlof},
  {Balsano}, {Bloch}, {Casperson}, {Fletcher}, {Gisler}, {Kehoe}, {Kinemuchi},
  {Lee}, {Marshall}, {McGowan}, {McKay}, {Rykoff}, {Smith}, {Szymanski}, \&
  {Wren}}]{wozniak+2004}
{Wo{\'z}niak}, P.~R., {Vestrand}, W.~T., {Akerlof}, C.~W., {et~al.} 2004, \aj,
  127, 2436

\bibitem[{{Wright} {et~al.}(2003){Wright}, {Egan}, {Kraemer}, \&
  {Price}}]{wright+2003}
{Wright}, C.~O., {Egan}, M.~P., {Kraemer}, K.~E., \& {Price}, S.~D. 2003, \aj,
  125, 359

\bibitem[{{York} {et~al.}(2000){York}, {Adelman}, {Anderson}, {Anderson},
  {Annis}, {Bahcall}, {Bakken}, {Barkhouser}, {Bastian}, {Berman}, {Boroski},
  {Bracker}, {Briegel}, {Briggs}, {Brinkmann}, {Brunner}, {Burles}, {Carey},
  {Carr}, {Castander}, {Chen}, {Colestock}, {Connolly}, {Crocker}, {Csabai},
  {Czarapata}, {Davis}, {Doi}, {Dombeck}, {Eisenstein}, {Ellman}, {Elms},
  {Evans}, {Fan}, {Federwitz}, {Fiscelli}, {Friedman}, {Frieman}, {Fukugita},
  {Gillespie}, {Gunn}, {Gurbani}, {de Haas}, {Haldeman}, {Harris}, {Hayes},
  {Heckman}, {Hennessy}, {Hindsley}, {Holm}, {Holmgren}, {Huang}, {Hull},
  {Husby}, {Ichikawa}, {Ichikawa}, {Ivezi{\'c}}, {Kent}, {Kim}, {Kinney},
  {Klaene}, {Kleinman}, {Kleinman}, {Knapp}, {Korienek}, {Kron}, {Kunszt},
  {Lamb}, {Lee}, {Leger}, {Limmongkol}, {Lindenmeyer}, {Long}, {Loomis},
  {Loveday}, {Lucinio}, {Lupton}, {MacKinnon}, {Mannery}, {Mantsch}, {Margon},
  {McGehee}, {McKay}, {Meiksin}, {Merelli}, {Monet}, {Munn}, {Narayanan},
  {Nash}, {Neilsen}, {Neswold}, {Newberg}, {Nichol}, {Nicinski}, {Nonino},
  {Okada}, {Okamura}, {Ostriker}, {Owen}, {Pauls}, {Peoples}, {Peterson},
  {Petravick}, {Pier}, {Pope}, {Pordes}, {Prosapio}, {Rechenmacher}, {Quinn},
  {Richards}, {Richmond}, {Rivetta}, {Rockosi}, {Ruthmansdorfer}, {Sandford},
  {Schlegel}, {Schneider}, {Sekiguchi}, {Sergey}, {Shimasaku}, {Siegmund},
  {Smee}, {Smith}, {Snedden}, {Stone}, {Stoughton}, {Strauss}, {Stubbs},
  {SubbaRao}, {Szalay}, {Szapudi}, {Szokoly}, {Thakar}, {Tremonti}, {Tucker},
  {Uomoto}, {Vanden Berk}, {Vogeley}, {Waddell}, {Wang}, {Watanabe},
  {Weinberg}, {Yanny}, \& {Yasuda}}]{york+2000}
{York}, D.~G., {Adelman}, J., {Anderson}, Jr., J.~E., {et~al.} 2000, \aj, 120,
  1579

\end{thebibliography}
\bibliographystyle{aa}

\onecolumn
\begin{longtable}{ccccccccc}\hline
\hline
\multicolumn {9}{c}{Properties of the stars in our sample}\\
\hline
\hline
\endhead
\hline
\hline
\multicolumn {9}{c}{Table continuing next page}
\endfoot
\multicolumn {9}{c}{}
\endlastfoot
\hline
GCVS name &  A.k.a. & Date & Ins. & $m_V$ & Var. & Spec. & Bin. &  Notes \\ \hline
\object{AI Pyx}&{\scriptsize GSC 07145-02569}&2006-04-01&H&6.23&IA&B3V&?&Old\,Spec\footnote{hereafeter OS}:\,B4V\\
\object{AL Leo}&{\scriptsize GSC 01415-01312}&2006-04-01&H&9.92&EA/D&F5V&Y&OS:\,F5\\
\object{AN Pyx}&{\scriptsize GSC 06039-01085}&2006-04-01&H&8.25&ACV&A2V-IV&?&OS:\,A0V\\
\object{AO Ant}&{\scriptsize GSC 07179-02338}&2006-06-11&F&8.56&?&M4III&VIS&OS:\,M0\\
\object{AQ Ant}&{\scriptsize GSC 07187-00952}&2006-06-13&F&9.02&?&M3III&?&OS:\,K5\\
\object{AR Ant}&{\scriptsize GSC 07708-00615}&2006-04-09&F&9.3&LB&M4III&?&OS:\,M\\
\object{AY Ant}&{\scriptsize GSC 07727-00703}&2006-06-12&F&9.84&?&K0e&?&OS:\,Kp\\
\object{BC Gru}&&2006-08-09&F&9.9&EW&K0+K0+K&SB3&OS:\,?, Old\,Bin\footnote{hereafter OB}:\,Y\\
\object{BH Cap}&{\scriptsize GSC 05762-02497}&2006-04-02&H&8.03&EB&F2IV+F2?&SB2&OS:\,F0, OB:\,Y\\
\object{BK Cap}&{\scriptsize GSC 06338-00216}&2006-05-12&F&8.78&LB&M1III&?&OS:\,M1/M2III:\\
\object{BM Scl}&{\scriptsize GSC 07512-00800}&2006-08-07&F&8.28&LB&M1III&?&OS:\,M2III\\
\object{BP Psc}&{\scriptsize GSC 05244-00148}&2006-06-29&H&9.04&IT&G2IVe&?&OS:\,?\\
\object{BP Scl}&{\scriptsize GSC 07512-00221}&2006-08-07&F&8.07&LB&K7III&?&OS:\,K5III\\
\object{BQ Scl}&{\scriptsize GSC 07507-01012}&2006-06-16&F&8.88&?&K5III&?&OS:\,M0III:\\
\object{BS Pav}&&2006-04-02&H&9.8&IS&M6&?&OS:\,?\\
\object{BW CMi}&{\scriptsize GSC 00206-01099}&2006-04-18&F&9.02&LB&K8V&?&OS:\,M0\\
\object{BW Pyx}&{\scriptsize GSC 06590-00111}&2006-04-18&F&9.93&SR&M2III&?&OS:\,M\\
\object{BZ Ind}&{\scriptsize GSC 09324-00534}&2006-08-10&F&8.72&LB&M4III&?&OS:\,M3III\\
\object{CL Phe}&{\scriptsize GSC 08457-00425}&2006-06-09&H&9.7&BY&K2IV-V&?&OS:\,K1V\\
\object{CP Cir} &{\scriptsize GSC 09019-01463}&2006-04-02&H&7.51&GCAS&B5V&?&OS:\,B5IV\\
\object{CS Gru}&{\scriptsize GSC 07996-00969}&2006-05-21&H&9.45&BY&G8V&Y&OS:\,G8/K0V, OB:\,?\\
\object{CX Gru}&{\scriptsize GSC 07997-00367}&2006-06-09&H&6.66&ELL&F8V+F8V&SB2&OS:\,F7V, OB:\,VIS\\
\object{DG Oct}&{\scriptsize GSC 09469-00374}&2006-06-13&F&8.90&?&M3III&?&OS:\,M3III:\\
\object{DG Psc}&{\scriptsize GSC 00575-00094}&2006-08-09&F&8.74&LB&M4III&?&OS:\,M\\
\object{DH Psc}&{\scriptsize GSC 00569-00289}&2006-08-09&F&8.69&LB&M3III&?&OS:\,K5\\
\object{DI Psc}&{\scriptsize GSC 00575-00918}&2006-08-08&F&7.28&LB&G8III-IV&?&OS:\,K0\\
\object{DK Psc}&{\scriptsize GSC 00576-00284}&2006-08-09&F&8.38&LB&M3III&?&OS:\,M\\
\object{DM Boo}&{\scriptsize GSC 00903-00938}&2006-08-12&F&8.73&BY&G2&?&Old\,Var\footnote{hereafter OV}:\,IB, OS:\,G5\\
\object{DR Cru}&{\scriptsize GSC 08655-01039}&2006-07-02&F&8.88&BY&K5V+K&SB2&OS:\,K3/K4V, OB:\,?\\
\object{DS Cha}&{\scriptsize GSC 09418-00570}&2006-04-13&F&8.96&LB&M2III&?&OS:\,M2\\
\object{DT Cha}&{\scriptsize GSC 09415-00568}&2006-04-21&F&8.56&LB&K3IV&?&OS:\,K3III\\
\object{DT Tuc}&{\scriptsize GSC 08839-00456}&2006-08-09&F&9.09&LB&M4III&?&OS:\,M3III\\
\object{DW Tuc}&{\scriptsize GSC 09130-00332}&2006-08-09&F&9.04&LB&M2III&?&OS:\,M1III\\
\object{DX Tuc}&{\scriptsize GSC 09130-01530}&2006-08-08&F&9.63&EW&F5+F&SB2&OS:\,F7IV/V, OB:\,Y\\
\object{EF Aqr}&{\scriptsize GSC 05248-01030}&2006-06-29&H&10.3?&ELL&G0+G?&SB2&OS:\,G0, OB:\,Y\\
\object{EO Boo}&{\scriptsize GSC 01481-00660}&2006-09-16&F&8.45&LB&M1III&?&OS:\,M2III\\
\object{FQ Leo}&{\scriptsize GSC 00267-00569}&2006-04-21&F&8.19&LB&M2IV-III&?&OS:\,K5\\
\object{FQ Lup}&{\scriptsize GSC 07339-01070}&2006-09-09&F&9.5&L&M7&?&OS:\,M5/M6II:\\
\object{GK Cnc}&{\scriptsize GSC 01401-00763}&2006-06-14&F&9.13&?&M4III&?&OS:\,M\\
\object{GSC 00244-00434}&&2006-04-18&F&10.4&?&F0IV&VIS&OS:\,F5\\
\object{GSC 7831-0069}&&2006-09-01&F&10.48&??&F0&?&OS:\,?\\
\object{GSC 9027-4849}&&2006-06-13&F&&?&K0+K?&SB2&OS:\,?, OB:\,?\\
\object{HD 109962}&{\scriptsize GSC 08232-01689}&2006-04-13&F&9.54&?&F0IV-III&?&OS:\,F2V\\
\object{HD 117316}&{\scriptsize GSC 09254-01886}&2006-04-13&F&8.23&?&F0III&?&OS:\,F2IV\\
\object{HD 156542}&{\scriptsize GSC 06241-00434}&2006-09-07&F&&?&F3IV-III&?&OS:\,F0V\\
\object{HD 89027}&{\scriptsize GSC 04910-01309}&2006-06-13&F&&?&F1V&?&OS:\,F0\\
\object{HD 95671}&{\scriptsize GSC 08619-02281}&2006-04-01&H&9.83&?&G0V&?&OS:\,G0\\
\object{HD 95673}&&2006-04-21&F&9.0&?&A0V&VIS&\\ 
\object{HX Lib}&{\scriptsize GSC 06752-00434}&2006-09-08&F&8.45&LB&M1III&?&OS:\,M0III\\
\object{HZ Lib}&{\scriptsize GSC 05580-00356}&2006-09-08&F&8.02&LB&M0III&?&OS:\,M\\
\object{IK Lib}&{\scriptsize GSC 06761-00434}&2006-09-08&F&7.91&LB&M1III&?&\\
\object{IM Vir}&{\scriptsize GSC 04955-00912}&2006-04-02&H&9.69&?&G8V+K2V&SB2&OS:\,G5, OB:\,Y\\
\object{IP Lib}&{\scriptsize GSC 06775-00038}&2006-04-02&H&9.97&BY&K0V&VIS&OS:\,G5/G6V\\
\object{IR Lup}&{\scriptsize GSC 08278-01151}&2006-09-01&F&8.49&LB&M2III&?&OS:\,M2/M3III\\
\object{IX Lib}&{\scriptsize GSC 05604-00579}&2006-09-08&F&8.54&LB&M2III&?&OS:\,M\\
\object{IX Lup}&{\scriptsize GSC 08698-00010}&2006-06-10&F&8.32&?&M2III&?&\\
\object{IX Vir}&{\scriptsize GSC 04931-00916}&2006-04-21&F&8.22&LB&M2III&?&OS:\,M\\
\object{IY Lib}&{\scriptsize GSC 05597-00200}&2006-09-08&F&7.9&LB&M0III&?&OS:\,K5\\
\object{IY Lup}&{\scriptsize GSC 07316-00591}&2006-09-08&F&8.23&LB&M1III&?&OS:\,M2III\\
\object{IY Vir}&{\scriptsize GSC 00865-00094}&2006-04-21&F&9.43&LB&K8III&?&OS:\,M0\\
\object{KK TrA}&&2006-08-11&F&9.5&L&M7&?&OS:\,M5Ib:\\
\object{KL Lup}&{\scriptsize GSC 07822-00250}&2006-09-09&F&8.23&LB&M0III&?&OS:\,K5III\\
\object{KM Aqr}&{\scriptsize GSC 05802-00227}&2006-08-10&F&8.13&LB&M3III&?&OS:\,M\\
\object{KM Lup}&{\scriptsize GSC 07830-02867}&2006-09-13&F&7.57&LB&M2&?&OS:\,M2III\\
\object{KN Com}&{\scriptsize GSC 00881-00011}&2006-06-13&F&8.81&?&M3III&?&OS:\,M0\\
\object{KN Lup}&{\scriptsize GSC 07313-00641}&2006-04-02&H&9.21&BY&G7V&SB2&OS:\,G0V, OB:\,?\\
\object{KN Vir}&{\scriptsize GSC 04944-01258}&2006-04-01&H&7.41&LB&K7III&?&OS:\,K5\\
\object{KO Lup}&{\scriptsize GSC 07317-00457}&2006-09-08&F&6.89&LB&M6III&?&OS:\,M3/M4III\\
\object{KP Lup}&{\scriptsize GSC 08299-02622}&2006-09-13&F&7.93&LB&M5III&?&OS:\,M4III\\
\object{KQ Aqr}&{\scriptsize GSC 06955-01363}&2006-06-04&H&9.53&ACV&A0V&?&\\
\object{KQ Vir}&{\scriptsize GSC 04941-00938}&2006-04-01&H&8.84&LB&M1III&?&OS:\,K5\\
\object{KS Aqr}&{\scriptsize GSC 05228-00455}&2006-08-10&F&9.42&LB&M4III&?&OS:\,M1\\
\object{KS Mus}&{\scriptsize GSC 09230-01722}&2006-04-13&F&8.07&LB&M1III&?&OS:\,M2III\\
\object{KX Lup}&{\scriptsize GSC 07837-01552}&2006-04-12&F&8.34&LB&M1III&?&OS:\,M2III\\
\object{KY Aqr}&{\scriptsize GSC 05239-01000}&2006-08-10&F&8.81&LB&M3III&?&OS:\,K5\\
\object{LO Aqr}&{\scriptsize GSC 05811-01661}&2006-08-10&F&7.44&IB&A9&?&OS:\,F0\\
\object{LO Mus}&{\scriptsize GSC 09000-00155}&2006-04-02&H&8.61&BY&K2V&?&OS:\,K0V\\
\object{LP Vir}&{\scriptsize GSC 05547-01517}&2006-04-01&H&6.92&ELL&F1IV&?&OS:\,F0\\
\object{LV Hya}&{\scriptsize GSC 07222-01221}&2006-04-01&H&6.2&ACV&A0III&?&OS:\,A0V\\
\object{LW Vir}&{\scriptsize GSC 05545-00751}&2006-04-09&F&9.14&LB&K8V-IV&?&OS:\,K5\\
\object{MS TrA}&{\scriptsize GSC 09041-00093}&2006-05-16&H&8.86&ACV&F2III&SB&OS:\,A9:IV:p, OB:\,?\\
\object{MS Vir}&{\scriptsize GSC 06141-00265}&2006-09-07&F&9.4&EW&K0+K5&SB2&OS:\,K0/K1III/IV, OB:\,Y\\
\object{MT TrA}&{\scriptsize GSC 09278-02611}&2006-08-27&F&8.44&LB&M1III&?&OS:\,M2III\\
\object{MW Vir}&{\scriptsize GSC 06147-00662}&2006-05-21&F&6.95&?&F0V&Y&OS:\,A5IV\\
\object{MY Vir}&{\scriptsize GSC 00320-00214}&2006-06-13&F&8.30&?&M3III&?&OS:\,M\\
\object{NN Hya}&{\scriptsize GSC 05451-00040}&2006-06-10&F&6.59&?&M0III&?&OS:\,K5\\
\object{NP Del}&{\scriptsize GSC 01647-00196}&2006-05-04&H&8.89&ELL&A3IV&?&OS:\,A0\\
\object{NP Hya}&{\scriptsize GSC 00229-00560}&2006-04-01&H&7.08&ACV&A0III&VIS&OS:\,A2\\
\object{NQ Hya}&{\scriptsize GSC 06031-00276}&2006-04-09&F&7.97&LB&K7IV&?&OS:\,M1III\\
\object{NQ Vel}&{\scriptsize GSC 07673-01114}&2006-04-01&H&7.63&IA&B5I+B8?&SB2&OS:\,A5, OB:\,?\\
\object{NR Peg}&{\scriptsize GSC 01649-00371}&2006-05-21&H&8.13&BY&G5III+?&SB2&OV:\,EB, OS:\,G0, OB:\,Y\\
\object{NR Vel}&{\scriptsize GSC 08158-01896}&2006-04-01&H&7.56&GCAS&B2e&?&OS:\,B2V:e\\
\object{NS Peg}&{\scriptsize GSC 01654-00963}&2006-08-06&F&7.92&LB&M3&?&OS:\,M\\
\object{NS Vel}&{\scriptsize GSC 07662-03038}&2006-06-06&F&7.32&?&B6IIIe&?&OS:\,B6III/IV\\
\object{NV Aps}&{\scriptsize GSC 09266-02415}&2006-05-21&F&8.87&?&K7III&?&OS:\,K5\\
\object{NV Hya}&{\scriptsize GSC 05462-01496}&2006-06-12&F&7.53&?&G8III&?&OS:\,K0\\
\object{NW Hya}&{\scriptsize GSC 04892-00755}&2006-06-14&F&7.75&?&K5III&?&OS:\,K5\\
\object{NX Peg}&{\scriptsize GSC 00550-00385}&2006-08-08&F&8.1&LB&M3III&?&OS:\,M\\
\object{OP Ser}&{\scriptsize GSC 00920-00630}&2006-06-10&F&8.32&?&K0III+?&SB2&OS:\,K0, OB:\,?\\
\object{OQ Hya}&{\scriptsize GSC 06034-00871}&2006-06-16&F&7.98&?&M0III&?&OS:\,M1III\\
\object{OS Peg}&{\scriptsize GSC 01122-00360}&2006-08-11&F&9.06&LB&M2III&?&OS:\,M0\\
\object{OT Peg}&{\scriptsize GSC 01679-00700}&2006-05-21&H&9.89&BY&G8IV+G2V&SB2&OS:\,K0, OB:\,?\\
\object{OU Aps}&{\scriptsize GSC 09286-01523}&2006-04-01&H&8.6&ACV&A1IV&?&OS:\,A0IV/V\\
\object{OU Hya}&{\scriptsize GSC 00241-01908}&2006-06-13&F&9.60&?&M3III&?&OS:\,M2\\
\object{OV Aps}&{\scriptsize GSC 09443-01367}&2006-04-01&H&8.14&ACV&A7V+A7V&SB2&OS:\,A7III, OB:\,?\\
\object{OV Hya}&{\scriptsize GSC 06038-00531}&2006-06-11&F&8.76&?&M4III&?&OS:\,M3/M4III\\
\object{OX Vel}&{\scriptsize GSC 08582-02734}&2006-04-01&H&7.6&ACV&F3IV-III&?&OS:\,A4m\\
\object{PP Hya}&{\scriptsize GSC 06054-00244}&2006-04-01&H&6.84&ELL&A5V&?&OS:\,A3III\\
\object{PP Vel}&{\scriptsize GSC 08582-00358}&2006-06-10&F&8.35&?&M1III&?&OS:\,M2/M3\\
\object{PQ Hya}&{\scriptsize GSC 06046-0017}&2006-06-11&F&9.00&?&M1III&?&OS:\,M0\\
\object{PT Hya}&{\scriptsize GSC 05489-01121}&2006-06-12&F&8.04&?&M2III&?&\\
\object{PU Hya}&{\scriptsize GSC 05496-00657}&2006-06-14&F&8.87&?&M4III&?&OS:\,M3III\\
\object{PW Peg}&{\scriptsize GSC 00566-01526}&2006-08-10&F&8.61&LB&M1III&?&OS:\,K5\\
\object{PW Ser}&{\scriptsize GSC 00367-00383}&2006-04-12&F&8.26&LB&M0III&?&OS:\,K5\\
\object{PW Vel}&{\scriptsize GSC 07678-02187}&2006-06-11&F&8.52&?&M5III&?&OS:\,M3III\\
\object{PX Peg}&{\scriptsize GSC 01695-01012}&2006-08-10&F&8.78&LB&M0III&?&OS:\,K5\\
\object{PX Ser}&{\scriptsize GSC 00379-00580}&2006-04-12&F&9.33&LB:&K2V+K5(?)&SB2&OS:\,K2, OB:\,?\\
\object{PY Hya}&{\scriptsize GSC 05505-00401}&2006-06-13&F&8.88&?&M2III&?&\\
\object{PZ Peg}&{\scriptsize GSC 01157-00471}&2006-08-10&F&9.65&LB&K0&?&\\
\object{QQ Hya}&{\scriptsize GSC 07186-00100}&2006-06-14&F&6.97&?&M2II-III&?&OS:\,M2III\\
\object{QQ Ser}&{\scriptsize GSC 05683-01296}&2006-05-24&F&7.66&?&M3III&?&OS:\,M2III\\
\object{QS Peg}&{\scriptsize GSC 01158-01116}&2006-08-10&F&7.96&LB&M0III&?&OS:\,K5\\
\object{QV Hya}&{\scriptsize GSC 07221-00698}&2006-04-09&F&9.05&LB&M0IV-III&?&OS:\,M1III:\\
\object{RX Gru}&{\scriptsize GSC 08008-00397}&2006-06-09&H&10.57&EB&G8V+G?&SB2&OS:\,?, OB:\,Y\\
\object{SX Equ}&{\scriptsize GSC 01103-02568}&2006-06-13&F&8.86&?&M1IV-III&VIS&OS:\,K5\\
\object{TV Sex}&{\scriptsize GSC 00239-01070}&2006-06-13&F&8.78&?&M2III&?&OS:\,K5\\
\object{TW Sex}&{\scriptsize GSC 04896-01383}&2006-06-13&F&7.93&?&M4III&?&OS:\,M4\\
\object{TYC 7798- 500-1}&&2006-05-21&F&&?&F5III&?&OS:\,?\\
\object{UV Crv}&{\scriptsize GSC 06108-00927}&2006-04-01&H&9.38&BY&K3IV+?&SB3&OS:\,K1V, OB:\,VIS\\
\object{UW Crt}&{\scriptsize GSC 06090-01429}&2006-04-09&F&8.24&LB&K5/7V-IV&?&OS:\,K5/M0III\\
\object{UW Sex}&{\scriptsize GSC 05495-00576}&2006-04-01&H&9.17&LB&M3III&VIS&OS:\,M0\\
\object{V1001 Cen}&{\scriptsize GSC 08682-01015}&2006-04-02&H&7.24&IA&B4&SB2&OS:\,B4IV/V, OB:\,?\\
\object{V1003 Sco}&{\scriptsize HD 149711}&2006-05-04&H&5.83&ELL&B3IV&VIS&OS:\,B2.5IV\\
\object{V1011 Cen}&{\scriptsize GSC 09011-04767}&2006-09-01&F&8.47&LB&M5&?&OS:\,M4/M5\\
\object{V1026 Sco}&{\scriptsize GSC 06199-00618}&2006-04-02&H&8.85&IA&F0IV-III&?&OS:\,A8Ve\\
\object{V1045 Sco}&{\scriptsize GSC 05624-00995}&2006-08-11&F&8.08&LB&K7III&?&OS:\,K5III\\
\object{V1048 Sco}&{\scriptsize GSC 07342-00752}&2006-08-11&F&9.05&LB&K7&?&OS:\,K5III:\\
\object{V1052 Sco}&{\scriptsize GSC 05613-00357}&2006-09-13&F&8.62&LB&M2&?&OS:\,K5\\
\object{V1053 Sco}&{\scriptsize GSC 05625-00721}&2006-04-12&F&8.03&LB:&M4IV-III&?&OS:\,M4III\\
\object{V1066 Sco}&{\scriptsize GSC 07358-00675}&2006-06-11&F&9.08&?&M2III&?&OS:\,M\\
\object{V1080 Sco}&{\scriptsize GSC 07883-00697}&2006-04-01&H&7.65&IA&A0V+A0V&SB2&OS:\,B9.5IV, OB:\,VIS\\
\object{V1085 Sco}&{\scriptsize GSC 07388-00568}&2006-05-24&F&9.02&?&M5III&?&OS:\,M2\\
\object{V1434 Aql}&{\scriptsize GSC 01030-00103}&2006-08-15&F&7.56&LB&M3III&?&OS:\,M\\
\object{V1440 Aql}&{\scriptsize GSC 05140-00148}&2006-05-18&F&8.4&ELL&B1V+B1V&SB2&OS:\,B8, OB:\,VIS\\
\object{V1442 Aql}&{\scriptsize GSC 01048-00621}&2006-05-18&F&7.46&LB&K7III-IV&?&OS:\,K0\\
\object{V1443 Aql}&{\scriptsize GSC 01040-00260}&2006-04-02&H&8.95&GCAS&B5Ve&?&OS:\,B9V\\
\object{V1450 Aql}&{\scriptsize GSC 01042-02041}&2006-04-02&H&8.98&EB&A0V+A?&SB2&OS:\,A0, OB:\,Y\\
\object{V1453 Aql}&{\scriptsize GSC 05143-00240}&2006-05-12&F&9.1&LB&M0IV&?&OS:\,K5\\
\object{V1459 Aql}&{\scriptsize GSC 01060-00614}&2006-08-08&F&8.48&LB&M3III+M&SB2&OS:\,M2, OB:\,?\\
\object{V1460 Aql}&{\scriptsize GSC 05728-01515}&2006-06-13&F&9.23&?&M2III&?&OS:\,K5\\
\object{V1465 Aql}&{\scriptsize GSC 00492-01519}&2006-05-12&F&9.22&ACV&F2V&?&OS:\,A5\\
\object{V1471 Aql}&{\scriptsize GSC 00498-01575}&2006-05-04&H&8.42&EB&A1V+A1V&SB2&OS:\,A0, OB:\,Y\\
\object{V2300 Oph}&{\scriptsize GSC 00442-01788}&2006-05-04&H&6.7&ELL&A0V&?&OS:\,A0\\
\object{V2350 Oph}&{\scriptsize GSC 05047-00021}&2006-09-13&F&7.41&LB&M3III&?&OS:\,M\\
\object{V2354 Oph}&{\scriptsize GSC 06221-00890}&2006-09-09&F&8.58&LB&M2III&?&OS:\,M2III:\\
\object{V2356 Oph}&{\scriptsize GSC 00979-00970}&2006-08-23&F&7.07&LB&G8I-II&?&OS:\,K5\\
\object{V2361 Oph}&{\scriptsize GSC 00410-02537}&2006-09-01&F&8.62&LB&M5&?&OS:\,M4\\
\object{V2362 Oph}&{\scriptsize GSC 00406-02288}&2006-09-01&F&8.72&LB&M5III&?&OS:\,M\\
\object{V2366 Oph}&{\scriptsize GSC 00399-01658}&2006-09-13&F&8.97&LB&M0&?&OS:\,K5\\
\object{V2369 Oph}&{\scriptsize GSC 00982-02178}&2006-04-01&H&8.54&BY&G9+?&SB3&OS:\,G5, OB:\,VIS\\
\object{V2370 Oph}&{\scriptsize GSC 00995-02245}&2006-09-01&F&9.79&LB&M0&?&\\
\object{V2386 Oph}&{\scriptsize GSC 00428-01318}&2006-09-01&F&7.41&LB&M4IV-III&?&OS:\,M\\
\object{V341 Sge}&{\scriptsize GSC 01606-01704}&2006-05-04&H&7.67&GCAS&B2Ve&?&OS:\,B2.5V\\
\object{V344 Hya}&{\scriptsize GSC 06710-00436}&2006-04-09&F&7.28&LB&M2III&?&\\
\object{V344 Peg}&{\scriptsize GSC 01712-00611}&2006-08-07&F&9.63&LB&M7III&?&OS:\,M5\\
\object{V349 Vel}&{\scriptsize GSC 08192-04033}&2006-04-01&H&9.64&ACV&F2+?&Y&OS:\,F3IVp\\
\object{V350 Nor}&{\scriptsize GSC 09036-03305}&2006-04-01&H&9.2&*&B5+?&SB2&OS:\,B8/B9Ib/II, OB:\,?\\
\object{V355 Hya}&{\scriptsize GSC 06744-00025}&2006-09-07&F&8.33&LB&M2III&?&\\
\object{V355 Pav}&{\scriptsize GSC 09290-00176}&2006-05-24&F&8.37&?&M2IV-III&?&OS:\,M2III:\\
\object{V357 Pav}&{\scriptsize GSC 09291-01224}&2006-05-04&H&7.94&ACV&B9III&?&OS:\,B8III\\
\object{V357 Vel}&{\scriptsize GSC 08615-02755}&2006-06-10&F&8.31&?&K0IV-III&?&OS:\,K0III\\
\object{V363 Nor}&{\scriptsize GSC 08710-01594}&2006-08-11&F&7.89&LB&M3III&?&OS:\,M2/M3III:\\
\object{V369 Pav}&{\scriptsize GSC 09313-00616}&2006-05-24&F&8.40&?&M2III&?&\\
\object{V370 Nor}&{\scriptsize GSC 08711-01144}&2006-08-27&F&8.17&LB&K7+?&SB2&OS:\,K5/M0III:, OB:\,VIS\\
\object{V371 Nor}&{\scriptsize GSC 08715-01929}&2006-04-02&H&9.35&BY&K5V&?&OS:\,K2V\\
\object{V375 Nor}&{\scriptsize GSC 08316-01234}&2006-09-13&F&9.79&LB&K2IV-III&?&OS:\,K5\\
\object{V384 Pav}&{\scriptsize GSC 09112-00897}&2006-05-12&F&7.99&LB&M3III&?&OS:\,M3/M4III\\
\object{V389 Pav}&{\scriptsize GSC 09321-01375}&2006-04-30&F&7.61&LB&M0IV&?&OS:\,M1/M2III\\
\object{V390 Pav}&{\scriptsize GSC 09114-01109}&2006-05-04&H&9.03&BY&K3V&VIS&OS:\,K2Vp\\
\object{V412 Pup}&{\scriptsize GSC 07659-03359}&2006-04-09&F&8.74&LB&M4III&?&OS:\,K5\\
\object{V414 Pup}&{\scriptsize GSC 05420-00844}&2006-04-01&H&8.80&ACV&A0I+A&SB2&OS:\,Ap, OB:\,Y\\
\object{V422 Pup}&{\scriptsize GSC 07133-02165}&2006-04-09&F&9.04&?&M4III&VIS&OS:\,K2\\
\object{V424 Pup}&{\scriptsize GSC 06563-02968}&2006-04-09&F&8.93&LB&K7IV-III&?&OS:\,K5\\
\object{V429 Pup}&{\scriptsize GSC 06000-01200}&2006-06-06&F&8.90&?&K5III&?&OS:\,K5\\
\object{V432 Pup}&{\scriptsize GSC 06005-05152}&2006-04-01&H&6.67&ACV&A3IV-V+A?&SB2&OS:\,A1IV, OB:\,?\\
\object{V435 Pup}&{\scriptsize GSC 07130-01521}&2006-04-09&F&8.33&LB&M0III&?&OS:\,K5\\
\object{V4376 Sgr}&{\scriptsize GSC 06258-00907}&2006-04-02&H&9.14&ELL&F8IV-III&?&OS:\,F7V\\
\object{V4377 Sgr}&{\scriptsize GSC 07399-00071}&2006-05-24&F&8.88&?&K7III&?&OS:\,K5\\
\object{V4385 Sgr}&{\scriptsize GSC 06855-03418}&2006-05-04&H&7.42&ELL&B8+F2p&SB2&OS:\,A2/A3V+, OB:\,?\\
\object{V4393 Sgr}&{\scriptsize GSC 06277-01745}&2006-05-24&F&7.63&?&K5+?&SB2&OS:\,K5/M0III:, OB:\,?\\
\object{V4413 Sgr}&{\scriptsize GSC 06883-00930}&2006-06-12&F&8.81&?&M4/5II&?&OS:\,M2/M3III\\
\object{V4422 Sgr}&{\scriptsize GSC 06312-00760}&2006-04-30&F&8.01&LB&M2IV&?&OS:\,M2III\\
\object{V4432 Sgr}&{\scriptsize GSC 07453-00754}&2006-06-13&F&8.27&?&M4III&?&OS:\,M2/M3III\\
\object{V4436 Sgr}&{\scriptsize GSC 07948-01704}&2006-06-11&F&8.92&?&M3III&?&OS:\,M2III:\\
\object{V4440 Sgr}&{\scriptsize GSC 07949-00463}&2006-05-24&F&8.74&?&K5III&?&OS:\,K5\\
\object{V494 Car}&{\scriptsize GSC 09213-01493}&2006-04-09&F&9.24&LB&M6III&?&OS:\,M5\\
\object{V538 Car}&{\scriptsize GSC 08967-00342}&2006-04-21&F&7.64&LB&M6III-II&?&OS:\,M3\\
\object{V715 CrA}&{\scriptsize GSC 07901-00567}&2006-05-04&H&6.8&ACV&B5+A3?&SB2&OS:\,A0II/IIIp, OB:\,?\\
\object{V771 Mon}&{\scriptsize GSC 04847-02925}&2006-06-06&F&8.14&?&M2III&VIS&OS:\,M\\
\object{V830 Ara}&{\scriptsize GSC 08342-04690}&2006-04-02&H&8.11&GCAS&B1III-IIe&?&OS:\,B2Ib/IIpe\\
\object{V841 Ara}&{\scriptsize GSC 08720-01374}&2006-05-04&H&8.73&BY&K0V&VIS&\\
\object{V844 Ara}&{\scriptsize GSC 08327-01945}&2006-08-23&F&8.38&LB&K0IV-III&?&OS:\,K2/K3III\\
\object{V845 Ara}&{\scriptsize GSC 08331-01099}&2006-09-07&F&8.37&LB&M5&?&OS:\,M3/M4III\\
\object{V856 Ara}&{\scriptsize GSC 08739-02203}&2006-08-31&F&8.34&LB&K5III&?&OS:\,K4III:\\
\object{V857 Ara}&{\scriptsize GSC 09052-01117}&2006-04-01&H&9.59&BY&K2V&?&OS:\,G8/K0V\\
\object{V870 Ara}&{\scriptsize GSC 08751-01331}&2006-05-24&F&8.92&?&G0&SB2&OS:\,F8, OB:\,Y\\
\object{V907 Her}&{\scriptsize GSC 00961-01721}&2006-06-11&F&8.31&?&K7V&?&OS:\,K5\\
\object{V915 Her}&{\scriptsize GSC 01524-00686}&2006-08-23&F&8.32&LB&M2III&?&OS:\,M\\
\object{V939 Cen}&{\scriptsize GSC 07783-00935}&2006-07-02&F&8.25&LB&M4III&?&OS:\,M3III\\
\object{V940 Cen}&{\scriptsize GSC 08257-01308}&2006-04-02&H&9.59&BY&K0III&?&OS:\,G8/K0V\\
\object{V944 Cen}&{\scriptsize GSC 07772-01464}&2006-04-09&F&9.23&LB&M2III&?&OS:\,K8\\
\object{V974 Cen}&{\scriptsize GSC 07269-00319}&2006-04-01&H&7.67&ELL&F2V+F?&SB2&OS:\,F2, OB:\,VIS\\
\object{V976 Cen}&{\scriptsize GSC 08999-00906}&2006-04-01&H&9.38&ACV&F0III-II&?&OS:\,A7IV/V\\
\object{V981 Her}&{\scriptsize GSC 01554-01928}&2006-08-15&F&7.24&LB&M2III&?&OS:\,M\\
\object{V982 Cen}&{\scriptsize GSC 07794-00587}&2006-04-02&H&9.6&BY&K4V+K&SB2&OS:\,K2V, OB:\,VIS\\
\object{V988 Cen}&{\scriptsize GSC 07280-01160}&2006-04-02&H&9.77&BY&K0+K+?&SB3&OS:\,K0V, OB:\,?\\
\object{V999 Her}&{\scriptsize GSC 01574-01910}&2006-06-12&F&8.90&?&M2&?&OS:\,M0\\
\object{VX PsA}&{\scriptsize GSC 07492-00927}&2006-07-01&F&6.97&LB&M3III&?&\\
\object{WY PsA}&{\scriptsize GSC 07511-00060}&2006-08-09&F&8.71&LB&M3III&?&OS:\,M2III\\
\object{XZ Pyx}&{\scriptsize GSC 06569-02459}&2006-04-09&F&9.22&LB&M3IV-III&?&OS:\,M\\
\object{ZZ Pyx}&{\scriptsize GSC 07139-01937}&2006-04-18&F&8.63&LB&K3V&?&OS:\,K2\\
\hline
\caption{The VSOP stars of the first observing season.}
\label{table1}
\end{longtable}

\twocolumn
\end{document}